\documentclass[life,article]{Definitions/mdpi} 

\usepackage{amsmath,amssymb}
\usepackage{hyperref}
\usepackage{tabularx}

\usepackage[version=4]{mhchem}

\newcommand{\sed}[1]{#1}

\firstpage{1} 
\pubvolume{1}
\issuenum{1}
\articlenumber{0}
\pubyear{2023}
\copyrightyear{2023}
\Title{A Surface Hydrothermal Source of Nitriles and Isonitriles}

\Author{Paul B. Rimmer $^{1}$\orcidA{}, Oliver Shorttle $^{2,3}$\orcidB{}}

\AuthorNames{Paul B. Rimmer, Oliver Shorttle}

\address{$^{1}$ \quad Cavendish Laboratory, University of Cambridge, JJ Thomson Ave, Cambridge, CB3 0HE, United Kingdom;  pbr27@cam.ac.uk\\
$^{2}$ \quad Institute of Astronomy, University of Cambridge, United Kingdom; os258@cam.ac.uk\\
$^{3}$ \quad Department of Earth Sciences, University of Cambridge, United Kingdom.}

\abstract{Giant impacts can generate transient hydrogen-rich atmospheres, reducing atmospheric carbon. The reduced carbon will form hazes that rain out onto the surface and can become incorporated into the crust. Once heated, a large fraction of the carbon would be converted into graphite. The result is that local regions of the Hadean crust were plausibly saturated with graphite. We explore the consequences of such a crust for a prebiotic surface hydrothermal vent scenario.  We model a surface vent fed by nitrogen-rich volcanic gas from high-temperature magmas passing through graphite-saturated crust.  \texorpdfstring{\sed{We consider this occurring at pressures of $1$--$1000 \; {\rm bar}$ and temperatures of $1500$--$1700 \, {\rm ^{\circ}C}$.} The equilibrium with graphite purifies the left-over gas, resulting in substantial quantities of nitriles ($0.1 \%$ \ce{HCN} and $1 \, {\rm ppm}$ \ce{HC_3N}) and isonitriles ($0.01 \%$ \ce{HNC}) relevant for prebiotic chemistry. \sed{We use these results to predict gas-phase concentrations of methyl isonitrile of $\sim 1 \, {\rm ppm}$}. \sed{Methyl isocyanide} can participate in the non-enzymatic activation and ligation of the monomeric building blocks of life,}{} and surface, or shallow, hydrothermal environments provide its only known equilibrium geochemical source.}


\begin{document}

%%%%%%%%%%%%%%%%%%%%%%%%%%%%%%%%%%%%%%%%%%

\section{Introduction}
\label{sec:intro}

Because the synthetic chemistry at life's origins is a many-step process, a requirement for a prebiotic environment is clean productive chemistry. If the prebiotic environment is too \sed{diverse and complex (if the number of different reacting species is too large)}, then the chemical parameter space inhabited by a geochemical environment becomes large, and the desired products and intermediates are lost in a morass of many thousands of other molecules. \sed{This is what we mean by messy chemistry: chemistry of such diversity and complexity, that desired chemical products and behaviors are harder to realize \cite{Islam2017,Sutherland2017,Rimmer2018}.} 

The requirement for clean chemistry is related to the arithmetic demon: if a step-wise reaction does not provide products with high and selective yields, and does not have a way of purifying and preserving desired products, then as the fraction of useful product becomes the reactant for the next step subsequent yields become exponentially diluted. Chemical reaction yields of 10\% over a sequence of twenty steps will take a starting solution with reactants at high concentration ($1 \; {\rm M}$) to a final product with a concentration of less than a molecule per cubic centimeter of solution. At some point, this ceases to be chemistry and becomes homeopathy.

Prebiotic chemistry is more likely to be successful when it is clean and productive.  This places constraints on the environments in which successful prebiotic chemistry can occur.  A prebiotic environment that hosts clean chemistry is one that facilitates selective, high-yield chemical reactions to occur. A productive environment for prebiotic chemistry is one that facilitates the synthesis of complex organic molecules. These two environmental conditions are in tension with each other. The only conditions that have been experimentally demonstrated to be productive are reducing environments, and reducing environments tend to be messy.
We will keep this tension in mind as we consider one of the particularly promising chemical starting points for prebiotic synthesis: nitriles.

Nitriles feature prominently in the majority of prebiotic systems chemistry \cite{Ruiz2014,Benner2019}. They carry the same redox state of nitrogen found in biomolecules \cite{Sasselov2020}, and have remained a constant in the highly fruitful iterative discovery of geochemically plausible UV-driven prebiotic synthesis of life's monomeric building blocks \cite{Sasselov2020,Green2021}. Isonitrile chemistry, specifically chemistry involving methyl isocyanide, has been discovered to have astounding chemical properties of activating and ligating monomeric building blocks \cite{Mariani2018}, including nucleotides \cite{Liu2020}, phospholipids \cite{Bonfio2020} and amino acids \cite{Liu2020,Wu2021}.

Paradoxes often bear fruit in scientific exploration \cite{Benner2014}. In addition to clean, productive chemistry, ideal conditions for the prebiotic chemistry that forms nitriles admit several other paradoxes.
\begin{itemize}
  \item Access to ultraviolet light at wavelengths between $200$--$400 \, {\rm nm}$ \cite{Patel2015}; and,\\
  shielding of ultraviolet light at wavelengths between $200$--$400 \, {\rm nm}$ \cite{Mariani2018}.
  \item Near-freezing temperatures \cite{Sanchez1966,Miyakawa2002,Rimmer2018,Rimmer2021Timescales}; and,\\
  near-boiling temperatures \cite{Mansy2008}.
  \item Low pH \cite{Mariani2018b}; and,\\
  neutral to high pH \cite{Mariani2018b}.
  \item Water activity $\sim 1$, the conditions under which most nitrile-based prebiotic chemistry takes place; and,\\
  water activity $\ll 1$, required for phosphorylation of nucleosides, and helpful for other condensation reactions \cite{Hulsof1976}.
\end{itemize}
It is not possible for a single static environment to fulfill all of these conditions. These can be satisfied in principle by a dynamic and heterogeneous environment. We show that surface hydrothermal vents fed by gas from high-temperature magma on early Earth can qualify.

Surface hydrothermal vents are exposed to ultraviolet light where they are in contact with the atmosphere, and shielded from ultraviolet light at depth or in crevices. \sed{Hydrothermal systems like the glaciovolcanic hydrothermal vents on Iceland today have a wide range of temperatures, pH and chemistries, and can provide good analogues for anoxic systems in the past \cite{Cousins2013,Ilanko2019}.} For a single hydrothermal vent, temperatures can be near freezing at their surfaces and above boiling at greater depth. Fluid flow through small channels of rock with natural mineral buffers towards a surface in contact with \ce{CO2}, or phosphate-rich alkaline lake water \cite{Toner2020}, can provide a steep gradient from high to low pH. Most of the vent has water activity near unity, but the surface edges of the vent can dry or freeze, lowering the water activity.

Hydrothermal vents, whether shallow surface vents or underwater vents, have redox gradients generated by serpentinization and radiolysis \cite{Lin2005,Shibuya2015}. It is unlikely that these processes would have generated sufficiently reducing conditions for the generation of nitriles or isonitriles \cite{Lin2005,Klein2013,Rimmer2019Hydro}. Primordial abiotic kerogen could provide sufficient reducing power, but it is debatable whether significant amounts of kerogen was present in the crust or upper mantle before life \cite{Gold1980,Glasby2006}.

Even if reducing conditions are accessible, it is unlikely that they would result in selective chemistry \cite{Rimmer2019Hydro}. Often the choice is reducing conditions or chemical selection. Clean selective chemistry tends to be oxidizing, or at least neutral. Reducing chemistry often results in tar \cite{Kim2016}.

In this paper, we show that \sed{clean productive chemistry rich in nitriles and isonitriles} can be found together in surface hydrothermal vents because of graphitization. We present the surface vent scenario in Section \ref{sec:scenario}. We discuss the model used to predict the surface vent chemistry in Section \ref{sec:model} and show our results in Section \ref{sec:results}. Section \ref{sec:discussion} contains discussion and conclusions.

\section{The Scenario}
\label{sec:scenario}

\begin{figure}[H]
\centering
\includegraphics[width=\textwidth]{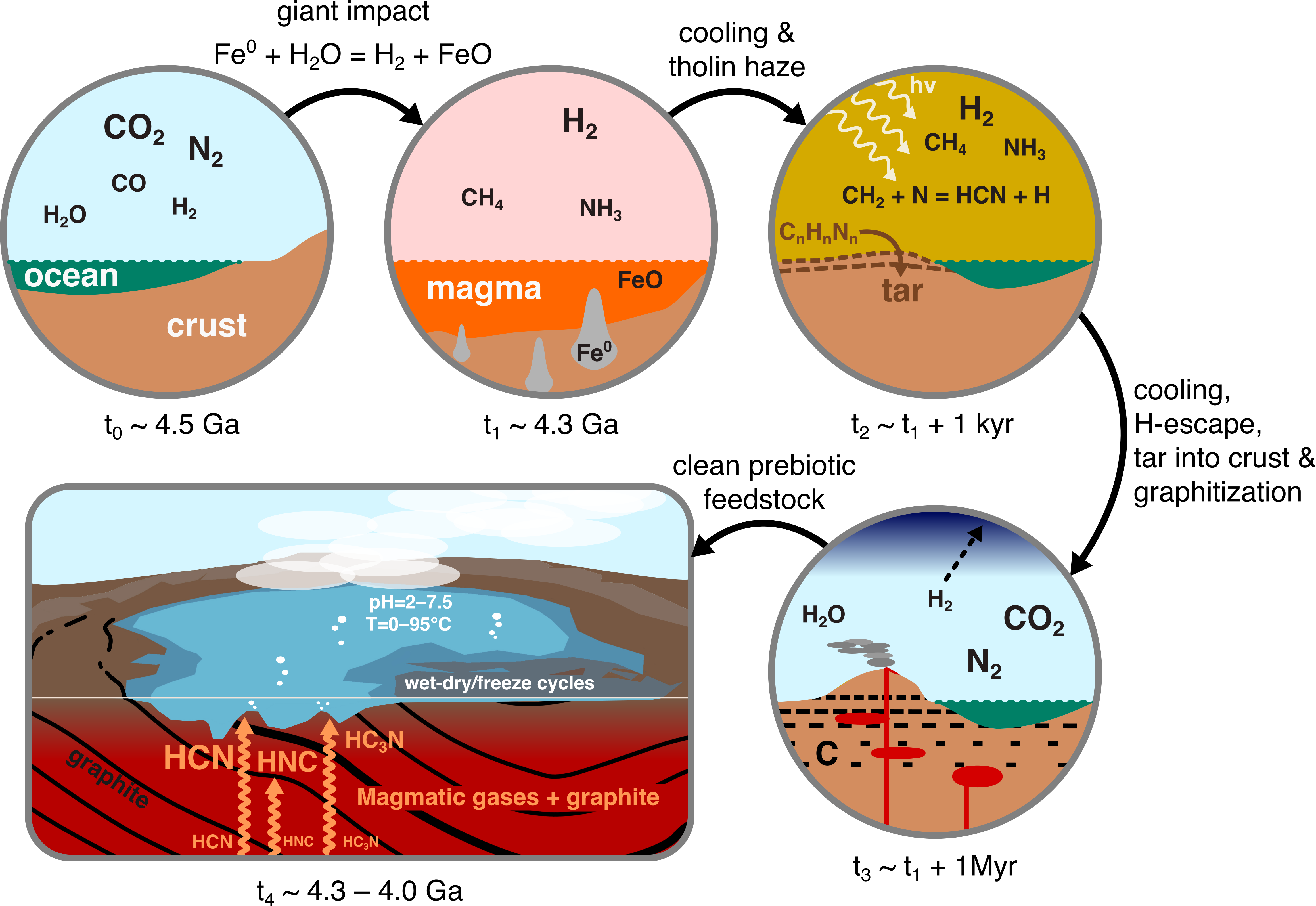}
\caption{A schematic representation of the scenario we propose here for clean high-yield production of prebiotic feedstock.  Events move around from top left clockwise: first Earth has a neutral atmosphere; this is reduced following giant impact at 4.3\,Ga by oxidation of the impactor's metal core  to produce a massive \ce{H2} atmosphere with significant methane and ammonia; This atmosphere quickly cools (in $<1$\,kyr) with photochemistry producing a tholin-rich haze that deposits complex nitrogen-rich organics; these organics become progressively buried and graphitized by interaction with magma, the atmosphere clears as \ce{H2} is lost to space and becomes neutral again; finally, magmatic gases interact with the graphite, and are scrubbed to produce high yields of clean HCN, \ce{HC3N} and isonitriles.   \label{fig:scenario}}
\end{figure}

We present a scenario that we predict to result in clean and productive prebiotic chemistry. It is a scenario that is both prebiotically plausible and well supported by observation, experiment and models. A schematic of this scenario is given in Figure \ref{fig:scenario}. We will discuss some of the simplifying assumptions of this scenario and other ways the same chemistry could emerge in Section \ref{sec:discussion}.

The Hadean Eon spans 500 million years of Earth's history, from planet formation to 4 Ga. After the moon-forming impact at $\sim 4.5 \, {\rm Ga}$, Earth likely had an atmosphere dominated by \ce{CO2} and \ce{N2}, with some \ce{CO} and \ce{H2O} \cite{Zahnle2010}, around $1-5 \%$ \ce{H2} \cite{Zahnle2018}, and comparatively trace amounts of sulfur-bearing compounds, \ce{SO2} and \ce{H2S} \cite{Ranjan2018}, with low-to-mid, stable concentrations of sulfites in most natural waters \cite{Ranjan2023}.

At a time around $\sim 4.3 \, {\rm Ga}$, Earth was likely hit by a roughly moon-sized object \cite{Genda2017}. The iron in this giant impactor would have reacted with ocean water, producing large amounts of hydrogen at high temperatures. Such a hydrogen-dominated atmosphere equilibrates with the surface magma generated to give $\sim$1\,bar partial pressure of \ce{H2} \cite{Itcovitz2022}.  In the high temperatures of this post-impact atmosphere, hydrogen would have reacted with carbon dioxide and nitrogen to produce methane and ammonia \cite{Zahnle2020,Wogan2023}. These giant impacts thereby initiated transient and global highly-reducing conditions in the atmosphere and on the crust of the Hadean Earth. Many of the reducing molecules, hydrogen, methane and ammonia, are greenhouse gases, and the surface of Earth at this time would have been hot, likely above the boiling point of liquid water at 1\,bar pressure \cite{Zahnle2020}.

In these conditions, the hydrogen, methane, and either nitrogen or ammonia in the atmosphere would have been photodissociated, with their products combining to produce the nitriles \ce{HCN} and \ce{HC3N} \cite{Wogan2023}.  Many complex organics would also have formed during this epoch, these would condense out of the atmosphere, forming a tholin-like haze \cite{Trainer2004,Arney2017,Zahnle2020}. This nitrogen-rich haze would have rained out onto the hot surface as a thick tar \cite{Benner2020,Ritson2020,Zahnle2020}. It is likely some of this tar was incorporated into the crust, either by tectonic, magmatic, or impact 
%(Anslow in prep. REF) 
churning of the surface. Over the period of a million years or more, the atmosphere would have returned to a neutral chemical state through the conversion of methane and ammonia back to carbon dioxide, nitrogen and hydrogen, and the escape of hydrogen into space \cite{Zahnle2020,Wogan2023}. The surface of Earth would then cool to near freezing \cite{Kadoya2020}.

Photochemically-produced tar mixed into the crust would have experienced episodic heating to $>1500 \, {\rm ^{\circ} C}$ by the early high-temperature (komatiitic) magmas known to have been an important constituent of the early Earth magmatism \cite{Nisbet1993}. We will show that this heating likely broke apart the tar, transforming most of it into graphite, molecular hydrogen and molecular nitrogen (though some of the hydrogen and nitrogen may have been complexed with the graphite at this stage). Magmatic gas, interacting with the graphitized crustal material, was likely transformed into \ce{HCN}, \ce{HC3N} and isonitriles, along with sulfide, carbon monoxide, and little else. The result is clean productive chemistry, which could degas through shallow and surface hydrothermal vents on ancient volcanic islands.

%%%%%%%%%%%%%%%%%%%%%%%%%%%%%%%%%%%%%%%%%%
\section{The Model}
\label{sec:model}

We will now model the latter part of this scenario: the interaction of gas initially in equilibrium with $f_{O2} \approx {\rm QFM-1}$ magma (i.e., one log unit below the quartz-magnetite-fayalite buffer in terms of its oxygen fugacity) as it flows through graphite. We start with a \sed{fiducial model at} $1700 \, {\rm ^{\circ}C}$, 100 bar gas at QFM$-1$ plus a fixed nitrogen content of $\ce{N} = 5.7\%$. Graphite is added to the system. \sed{QFM$-1$ conditions are incorporated by setting initial conditions to the values from Table \ref{tab:initial-conditions}.}

\sed{We are not here invoking nitrogen-rich magma. The nitrogen is included in the magmatic gas only for convenience; it makes no difference if it is included initially, or included with the graphite, because the quantity is not varied. The source of the nitrogen is expected to be the same as the source of the graphite. This nitrogen concentration represents the predicted nitrogen-rich nature of the post-impact organics \cite{Wogan2023}, that were then incorporated into the crust.} 

To predict the change in concentrations of chemical species as a function of added graphite, we solve the following equation,
\begin{equation}
\dfrac{d[\ce{X}]}{dt} = P_{\rm X} - L_{\rm X} \ce{[X]},
\end{equation}
where $[\ce{X}] \, {\rm (cm^{-3})}$ is the concentration of species \ce{X}, $P_{\rm X} \, {\rm (cm^{-3} \, s^{-1})}$ is the rate of production for that species, and $L_{\rm X} \, {\rm (s^{-1})}$ is the rate of loss for that species. The terms $P_{\rm X}$ and $L_{\rm X}$ are made up of concentrations of other species and rate constants, $k$, which themselves have units depending on the reaction order, and, when reversible, are set to reproduce chemical equilibrium by assuring that, for the generic reaction:
\begin{equation}
\ce{A} + \ce{B} + \ce{C} + ... \rightleftarrows \ce{X} + \ce{Y} + \ce{Z} + ...,
\end{equation}
with forward rate constant $k_+$ and reverse rate constant $k_-$, the reverse rate constant is set such that:
\begin{equation}
\dfrac{k_+}{k_-} = K_{\rm eq} = e^{-\Delta_f G/RT} = \dfrac{\{\ce{X}\}\{\ce{Y}\}\{\ce{Z}\}...}{\{\ce{A}\}\{\ce{B}\}\{\ce{C}\}...},
\end{equation}
where $\{\ce{A}\}$, etc. are the activities of the different chemical species, and $\Delta_f G/RT$ is determined using NASA coefficients, mostly from Burcat \& Ruscic \cite{Burcat2005}, and subsequent updates to the database. More details can be found in prior presentations of the underlying model \cite{Rimmer2016,Rimmer2021Venus}.

For these calculations, we use an updated gas-phase chemical network based on STAND-2020 \cite{Rimmer2021Venus}. \sed{This model includes H/C/N/O/S chemistry, a very limited P network, and some reactions involving various heavier elements such as Fe, Mg, Ti. It includes 6279 reactions involving 511 chemical species} The \sed{full} list of species and network are available at \url{https://doi.org/10.7910/DVN/FKKYY3}. The main addition is the two reactions:
\begin{equation}
\ce{C(g)} \rightleftarrows \ce{C(s)},
\end{equation}
where \ce{C(g)} is gas-phase carbon and \ce{C(s)} is solid carbon as graphite \sed{and the rate constants are set to reproduce equilibrium}. The model is run for one day model time. 

\sed{The kinetics model and the FastChem model use different vapor pressures for graphite. For the kinetics model we use \cite{Brewer1948,Joseph2002}:
\begin{equation}
\log_{10} \, p_{\rm vap} = 6.455 - \dfrac{2.7709 \times 10^4}{T - 3.549} - \dfrac{10^7}{T^2},
\label{eqn:high-vap}
\end{equation}
where $p \, ({\rm bar})$ is the pressure and $T \, ({\rm K})$ is the temperature. 

We want to compare our kinetics results to equilibrium. We predict the equilibrium of the magmatic gas using FastChem \cite{Stock2022}, to which we have added the thermochemical data for cyanoacetylene \citep[\ce{HC3N}; ][]{knight1985selected}, given as the constant of mass action
\begin{equation}
\ln{K(T)} = \dfrac{2.96\times10^5}{T} - 4.83\ln{T} - 28.47 + \big(1.91\times10^{-3}\big)\, T - \big(1.10\times10^{-7}\big) \, T^2,
\label{eqn:low-vap}
\end{equation}
where $T$ is the temperature in Kelvin. For FastChem the vapor pressure works out to be approximately:
\begin{equation}
\log_{10} \, p_{\rm vap, FC} = 4.855 - \dfrac{2.5709 \times 10^4}{T} - \dfrac{10^7}{T^2}.
\end{equation}

We run our model for a wide range of conditions to determine the sensitivity of our results to elemental composition, temperature and pressure. We run the model from $1300 \, {\rm ^{\circ} C}$ to $1800 \, {\rm ^{\circ} C}$ and from $1 \, {\rm bar}$ to $1000 \, {\rm bar}$, varying \ce{H}, \ce{C}, \ce{N} and \ce{O} elemental abundances.}

\sed{Our network does not include methyl isocyanide (\ce{CH3NC}), the prebiotically-relevant compound, and the gas-phase kinetics of this species would require more investigation before it could be reliably included in this model. We show \ce{HNC} as an indication of the overall concentration of isonitriles, and we use the BURCAT thermochemistry data to estimate the concentration of \ce{CH3NC}. 

We can use the Gibbs free energy $(\Delta_r G, \, {\rm kJ/mol})$ for the following reaction:
\begin{equation}
\ce{HCN} + \ce{CH4} \rightarrow \ce{CH3NC} + \ce{H2},
\label{eqn:CH3NC}
\end{equation}
to predict \ce{CH3NC} concentration. This can be expressed as:
\begin{equation}
[\ce{CH3NC}] = e^{-\Delta_r G/RT} \dfrac{[\ce{HCN}][\ce{CH4}]}{\ce{[H2]}}.
\label{eqn:CH3NC-equil}
\end{equation}
where [\ce{HCN}] and [\ce{CH4}] are the concentrations of \ce{HCN} and \ce{CH4}, respectively; $R = 8.3145 \; {\rm J/(mol \, K}$ is the gas constant, $T \, ({\rm K})$ is the temperature.}

\begin{table} 
\caption{\sed{Initial conditions for the kinetics model. We set temperature equal to $1700 \, {\rm ^{\circ}C}$ and pressure is $100 \, {\rm bar}$. Mixing ratios of species are given below. Abundances have been rounded. Exact abundances used can be found at \url{https://doi.org/10.7910/DVN/FKKYY3}.} \label{tab:initial-conditions}}
\begin{tabularx}{\textwidth}{ccccccccc}
\toprule
\textbf{Species} & \ce{C(g)} & \ce{CO} & \ce{H2} & \ce{N2} & \ce{O2} & \ce{CH4} & \ce{CO2} & \ce{H2O}\\
\midrule
\textbf{Mixing Ratio}	& \textsuperscript{*}	& $0.35$ & $0.1$ & $0.057$ & $7.6 \times 10^{-10}$ & $1.4 \times 10^{-6}$ & 0.23 & 0.26\\
\bottomrule
\end{tabularx}
\noindent{\footnotesize{\textsuperscript{*} Graphite content is varied from zero to saturation (elemental abundance $\approx$ 80\%)}}
\end{table}

\section{Results}
\label{sec:results}

\begin{figure}[H]
\centering
\includegraphics[width=0.8\textwidth]{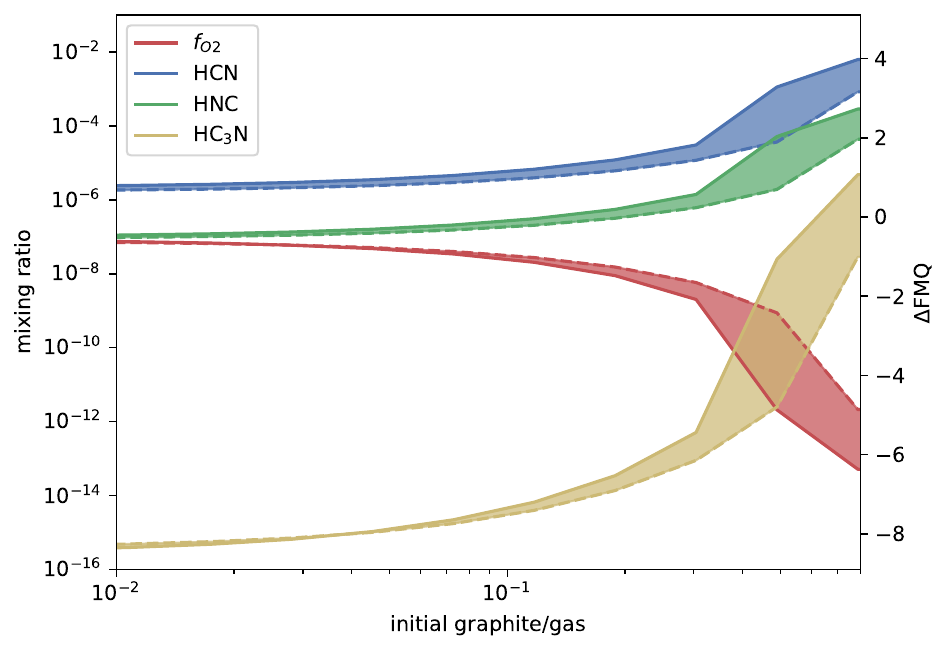}
\caption{Mixing ratios of \ce{HCN}, \ce{HNC} and \ce{HC3N} \sed{(left y-axis)} as a function of added graphite (x-axis) \sed{for a QFM-1 buffered magmatic gas with $5.7 \%$ elemental nitrogen content held at} $1700 \, {\rm ^{\circ}C}$ and $100 \, {\rm bar}$. \sed{The right y-axis shows the deviation from QFM in log units. Solid lines indicate the kinetics (this paper) results and dashed lines the equilibrium (FastChem) results, with the range between the two shaded in.}\label{fig:QFM-graphite}}
\end{figure}

\begin{figure}[H]
\centering
\includegraphics[width=0.8\textwidth]{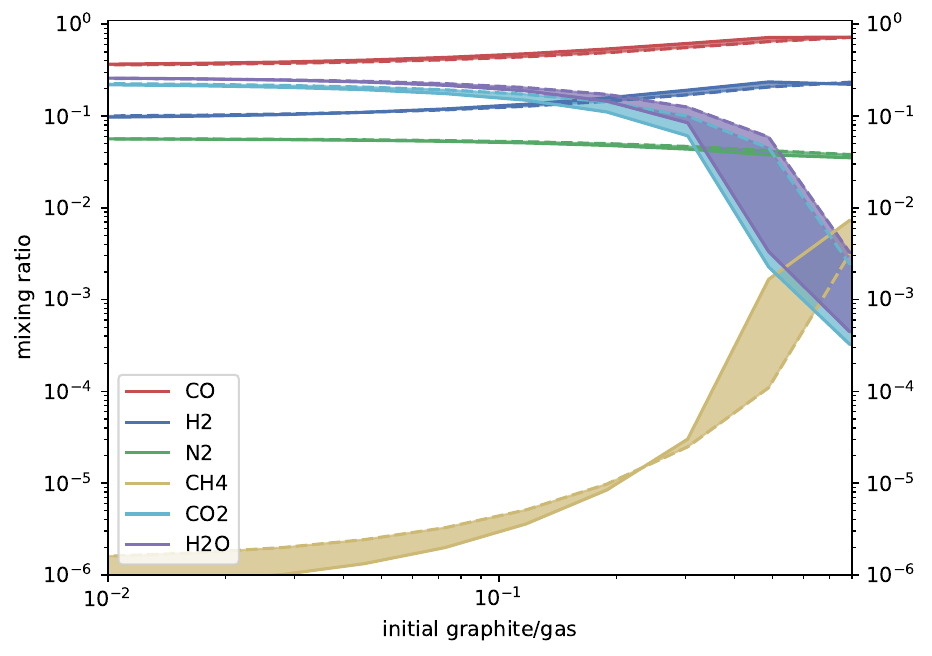}
\caption{\sed{Mixing ratios of major species (y-axis)  for a QFM-buffered magmatic gas with $5.7 \%$ elemental nitrogen content held at $1700 \, {\rm ^{\circ}C}$ and $100 \, {\rm bar}$. Solid lines indicate the kinetics (this paper) results and dashed lines the equilibrium (FastChem) results, with the range between the two shaded in.}\label{fig:QFM-graphite-major}}
\end{figure}

At $1700 \, {\rm ^{\circ}C}$ and 100 bar, the model achieves equilibrium after $10^3$ -- $10^4 \, {\rm s}$ when the graphite concentration is $\lesssim 0.1$. The kinetics and equilibrium models then diverge. The equilibrium model does not change with graphite added beyond a concentration of 0.8, which is the saturation limit of graphite. The kinetics model continues to deviate from equilibrium, but much more slowly. The results below saturation are given in Figure \ref{fig:QFM-graphite} \sed{for \ce{HCN}, \ce{HNC} and \ce{HC3N}, and Figure \ref{fig:QFM-graphite-major} for major species.} A datafile of the full results is available at \url{https://doi.org/10.7910/DVN/FKKYY3}. Results are given as a function of carbon and oxygen fractions. 

\begin{figure}[H]
\centering
\includegraphics[width=0.8\textwidth]{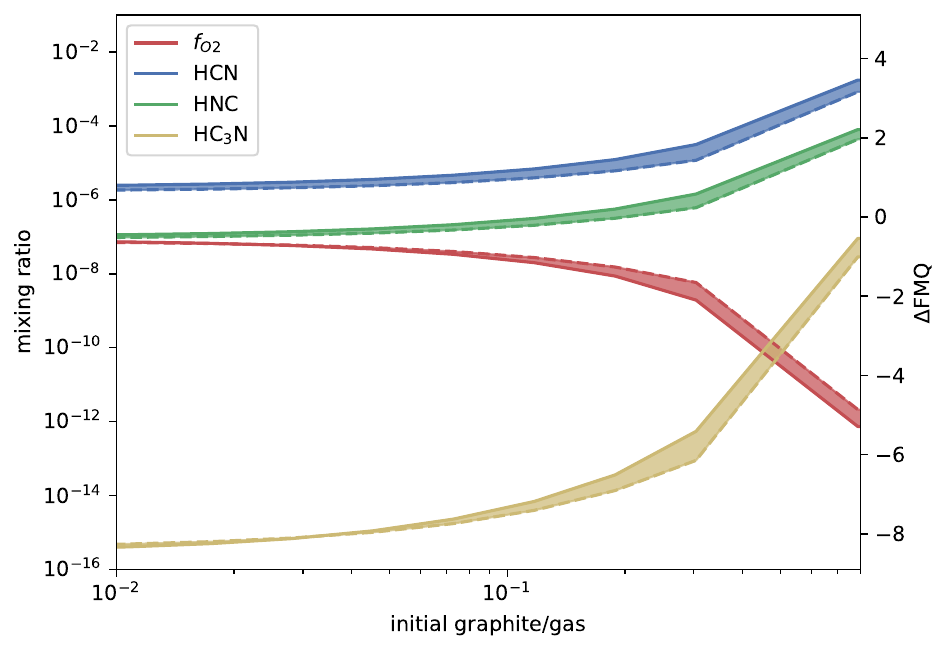}
\caption{\sed{Mixing ratios of \ce{HCN}, \ce{HNC} and \ce{HC3N} (left y-axis) as a function of added graphite (x-axis) for a QFM-buffered magmatic gas with $5.7 \%$ elemental nitrogen content held at $1700 \, {\rm ^{\circ}C}$ and $100 \, {\rm bar}$. Here the kinetics model and the equilibrium model both use graphite vapor pressure equal to Eq. (\ref{eqn:low-vap}). The right y-axis shows the deviation from QFM in log units. Solid lines indicate the kinetics (this paper) results and dashed lines the equilibrium (FastChem) results, with the range between the two shaded in.}\label{fig:QFM-conv}}
\end{figure}

\sed{The main reason for the deviation between the two models is the choice of graphite vapor pressure, where the kinetics model uses Eq. (\ref{eqn:high-vap}) and the equilibrium model uses Eq. (\ref{eqn:low-vap}). If the kinetic model is run with the graphite vapor pressure equal to Eq. (\ref{eqn:low-vap}), the results converge, as can be seen in Figure \ref{fig:QFM-conv}.}

\sed{We also show how the concentrations of \ce{HCN}, \ce{HNC} and \ce{HC3N} at graphite saturation depend on the magma temperature, see Figure \ref{fig:QFM-Temp}. The results of the sensitivity analysis are presented in the Appendix. This analysis was only run for the chemical kinetics model.}

\begin{figure}[H]
\centering
\includegraphics[width=0.85\textwidth]{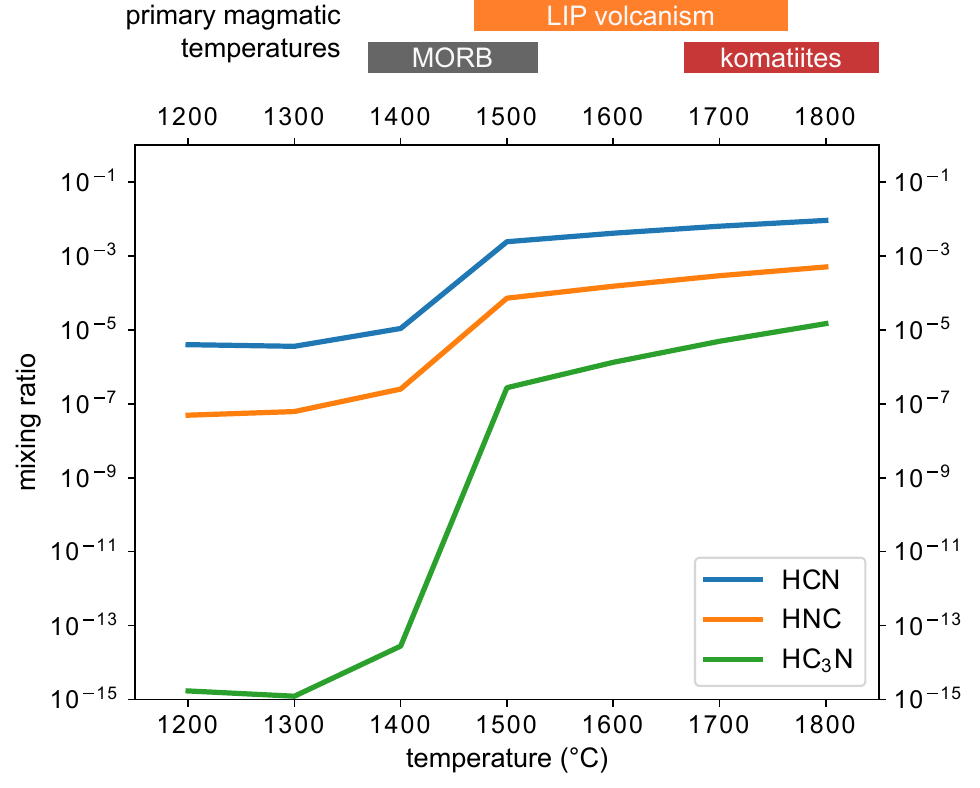}
\caption{\sed{Mixing ratios of \ce{HCN}, \ce{HNC} and \ce{HC3N} (y-axis) at graphite saturation as a function of temperature ($^\circ$C, x-axis) for a magmatic gas stating at QFM$-1$ with $5.7 \%$ elemental nitrogen content held at $100 \, {\rm bar}$. Only the kinetics results are shown.  Indicative eruptive temperatures of terrestrial magmas are given above the plot: `MORB' = mid-ocean ridge basalts \citep{coogan2014_chemgeol}, `LIP' = Large igneous province \citep{coogan2014_chemgeol}, and komatiite temperatures are given for Archean-age examples \citep{Nisbet1993}.}\label{fig:QFM-Temp}}
\end{figure}

The most important result is that graphite formation cleans up the chemistry considerably. While Rimmer \& Shorttle \cite{Rimmer2019Hydro} predict that the majority carbon-containing species is diacetylene, our study finds that the majority carbon-containing species is graphite, and diacetylene has decreased from $42 \%$ to $< 0.1\%$. This is significant because diacetylene at lower temperatures will polymerize and will plausibly generate a mass of large inert hydrocarbons, effective ``tarrification'' \cite{Kim2016}. Graphite formation resolves much of this problem. Since graphite is removed from the gas phase, the remaining therochemically stable gas is enriched, and so much higher concentrations of cyanide (\ce{HCN}) and cyanoacetylene (\ce{HC_3N}) are achieved, to a maximum of $\sim 1\%$ and $\sim 1 \, {\rm ppm}$ of the gas-phase respectively.

An unexpected prediction is the formation of higher-than-expected concentrations of hydrogen isocyanide (\ce{HNC}). At graphite saturation, $[\ce{HNC}] \sim 0.01\%$.

\begin{figure}[H]
\centering
\includegraphics[width=0.8\textwidth]{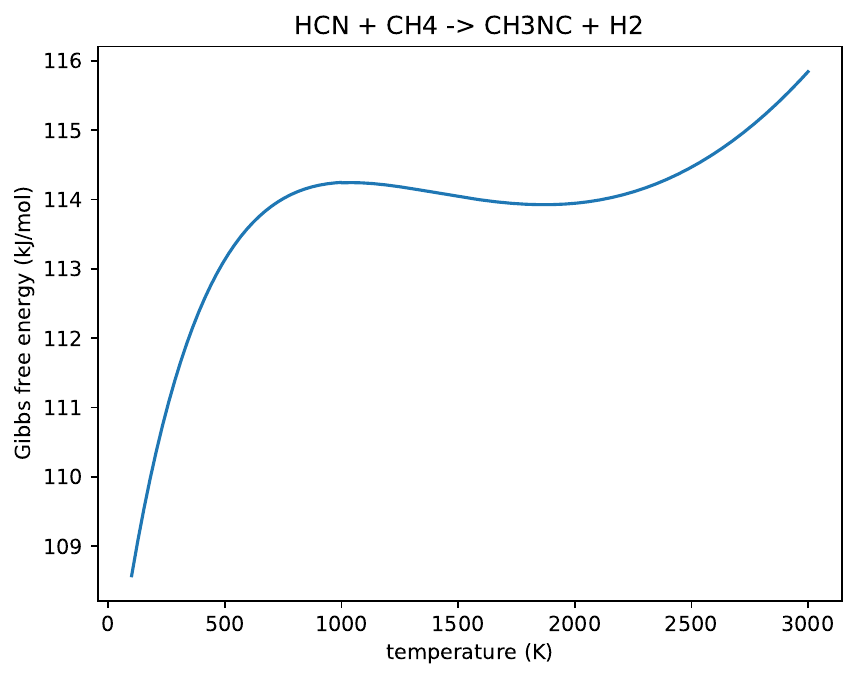}
\caption{\sed{The Gibbs free energy of the reaction: $\ce{HCN} + \ce{CH4} \rightarrow \ce{CH3NC} + \ce{H2}$ (kJ/mol), as a function of temperature (K). We can see that for all temperatures, this formation reaction for \ce{CH3NC} is endergonic, but not strongly so.}\label{fig:Gibbs-CH3NC}}
\end{figure}

\sed{We use the data from BURCAT to calculate the Gibbs free energy of Reaction (\ref{eqn:CH3NC}), and show this free energy in Figure \ref{fig:Gibbs-CH3NC}. We apply the value of the Gibbs free energy at $1700 \, {\rm ^{\circ}C}$ of $114 \, {\rm kJ/mol}$, and $[\ce{H2}] = 0.2$, $[\ce{CH4}] = 3-7 \times 10^{-3}$ and $[\ce{HCN}] = 10^{-3} - 10^{-2}$ to Equation (\ref{eqn:CH3NC-equil}) to predict the equilibrium concentration:
$[\ce{CH3NC}]$ is between $7 \times 10^{-7}$ and $2 \times 10^{-5}$.}

\sed{At lower pressure, if the magma is graphite saturated, then \ce{HCN}, \ce{HNC} and \ce{HC3N} remain effectively unchanged. In systems which are not graphite saturated, \ce{HCN} and, in the right range, \ce{HNC} become even more favored}, with \ce{HNC} achieving mixing ratios of $\sim 0.1\%$ of the gas phase. Cyanoacetylene drops significantly, by roughly two orders of magnitude per order of magnitude of decreasing pressure, to $\lesssim 100 \, {\rm ppt}$ levels at 1 bar. \sed{In any event, experiments to test the predictions of this model at ambient pressure, when the system is graphite saturated, may be tractable in the near future.}

At lower temperature, graphite reigns. Rimmer \& Shorttle \cite{Rimmer2019Hydro} predict that, at $1200 \, {\rm ^{\circ}C}$, significant amounts of \ce{HCN} and \ce{HC3N} could still be formed at $>10 \, {\rm bar}$ pressures. With our present model, the carbon is largely removed into graphite, and species like methane, molecular hydrogen and carbon monoxide dominate. For model results over a full range of oxygen, carbon and hydrogen fractions, see Appendix \ref{app:range}.   

\section{Discussion and Conclusions}
\label{sec:discussion}

In this paper, we present a new calculation of magmatic gas-phase chemistry for nitrogen-rich, otherwise of standard oxidation state (QFM$-1$), magmas at $1700 {\rm ^{\circ}C}$ temperatures and 100\,bar pressure, with graphite added to the point of saturation. This is the specific sequence we have modeled, but it may well turn out that the QFM$-1$ magma is nitrogen-poor and instead that at temperatures of $\sim 1700 {\rm ^{\circ}C}$ nitrogen is made available to the gas from the its complex with graphite. This model differs from Rimmer \& Shorttle 2019 \cite{Rimmer2019Hydro} in that we now include phase equilibrium with graphite. We find that graphite helps to clean up the chemistry, and increases the concentrations of prebiotically relevant compounds, nitriles and isonitriles, if the magma temperature is $\sim 1700 \, {\rm ^{\circ}C}$, consistent with early experimental studies of nitrogen and graphite at high temperatures \cite{Palmer1967}. \sed{If the magma temperature is lower, the concentration of nitriles and isonitriles drops significantly, see Figure \ref{fig:QFM-Temp}.}

\sed{Including graphite is not predicted to result in a decrease of nitriles so long as the temperature is sufficiently high ($\gtrsim 1500 {\rm ^{\circ}C}$). The results at lower temperatures indeed show a significant suppression of nitriles, in line with previous model expectations for hydrogen solubility, such as those of Wogan et al. \cite{Wogan2020}. It is important to note that these environments are locally at least an order of magnitude more reducing than the minimum $f_{\ce{O2}}$ considered by Wogan et al. Even under these conditions, the roughly estimated kg gas/kg magma for \ce{CH4} at 100 bar and 1700 K would be $\lesssim 10^{-4}$, in reasonable agreement with Wogan et al. If the magma was still present at the point of gas-graphite interaction, solubility experiments would need to be performed under these unusual conditions in order to determine the true fate of gas-phase nitriles and other hydrogen- and nitrogen-bearing species.}

\sed{The required conditions for clean chemistry with nitriles and isonitriles is graphite-saturated magma at $\gtrsim 1500 {\rm ^{\circ}C}$, else the carbon and nitrogen that would form nitriles is locked up in \ce{N2} and graphite. See the appendix for plots showing the dependence of the nitrile concentration on temperature and concentration of carbon (Figures \ref{fig:graphite-1300C} - \ref{fig:HNC-1300C}). Cyanoacetylene is favored over a wide range of parameter space at higher pressure ($\gtrsim 100 \, {\rm bar}$; see the Appendix, Figures \ref{fig:graphite-10bar} - \ref{fig:HNC-10bar}), but can we predict that it can be produced in abundance at pressures as low as $1 \, {\rm bar}$ so long as the system is saturated with graphite.  See also Table \ref{tab:range-summary} for a summary of these comparisons.}

The required high temperatures effectively limit the source of nitriles and isonitriles to more niche Hadean environments, particularly those that run at higher temperatures, such as Komatiite magmas \cite{Nisbet1993,herzberg2010thermal}, \sed{for which temperatures can surpass $1600 \, ^{\circ}C$ \cite{Takahashi1985}.} These high-temperature magmas are thought to be more prevalent (though not ubiquitous) on the Hadean and Archaean Earth \cite{Campbell1989,Nebel2014}. This development of the theory admits to several potential scenarios beyond the one we presented in Section \ref{sec:scenario}.

It may simply be that certain regions of the early upper mantle were highly reduced, possessing abiotically-generated kerogen-like material, in a manner hypothesized by Thomas Gold, among others \cite{Gold1980}. The natural high temperatures of early volcanic systems could heat and reprocess this material.

Cosmic dust is kerogen-like, and would have been much more ubiquitous on early Earth, with as much as 50\% of the dust being cosmic during the Hadean \cite{Walton2023}. The chemistry predicted here could arise in environments where this dust is concentrated and then heated volcanically, or by a subsequent large impact.

Radiolysis and serpentinization can generate large redox gradients \cite{Lin2005,Rempfert2023}, with certain regions of the crust becoming much more reduced while others become more oxidized. If the reduced regions intersected a magmatic flow at $>10 \, {\rm bar}$, the heat would convert much of the chemistry to nitriles and isonitriles. Radiolysis especially would have been much more intense during the Hadean \cite{Pastorek2020}.

Though we favor the scenario presented in Section \ref{sec:scenario} for Early Earth, based on current geological and experimental evidence, we are encouraged that similar prebiotic chemistry can emerge in a wide range of conditions expected on early Earth, Mars, and on exoplanets. It is also worth noting that all of these scenarios are compatible with shallow hydrothermal systems, which admit most of the physical and chemical advantages of underwater hydrothermal events \cite{Barge2022}. \sed{It is also worthy of note that some of these systems, such as the surface hydrothermal fields near Erebus volcano, Antarctica, exhibit these vast temperature shifts, with high-temperature magmatic systems intersecting with vents that release gas into ice \cite{Ilanko2019}, a system that would favor concentrating some of these prebiotic feed-stocks into eutectic phases. In addition, these systems show fascinating possibly-abiotic redox behavior, and can incorporate nitrogen from the atmosphere into their magmatic systems \cite{Ilanko2019}.}

The provision of chemical feedstocks from \sed{high-temperature} magmas may pave the way for order amidst geochemical chaos, with clean chemical equilibrium mixtures at high-concentration, segregated by pool or stream on the basis of the magmatic source: magmas with higher carbon content provide isonitriles, \sed{such as methyl isocyanide}. All of this is mediated and regulated by the formation of graphite, removing from the gas excess carbon and the combinatorial mess that comes with it, and high temperatures, that favor relatively simple gas-phase mixtures of the starting material required for productive prebiotic chemistry. Natural prebiotic environments need not produce messy chemistry; \sed{the environment can constrain the chemistry to be clean and productive}. Even if the environments favouring more chemically ordered and promising prebiotic chemistry turn out to be a very small fraction of the total environment, the selection pressure for productive synthesis could outweigh their relative rarity.

Whether this chemical solution provides a ``buffet lunch'' for prebiotic chemistry, or an unappetizing and unusable mess depends on the kinetic stability and solubility of the gas-phase mixture once it is quenched and enters into the surface waters. This question can only be resolved with future experiments.

%%%%%%%%%%%%%%%%%%%%%%%%%%%%%%%%%%%%%%%%%%
\vspace{6pt} 

%%%%%%%%%%%%%%%%%%%%%%%%%%%%%%%%%%%%%%%%%%

%\authorcontributions{For research articles with several authors, a short paragraph specifying their individual contributions must be provided. The following statements should be used ``Conceptualization, X.X. and Y.Y.; methodology, X.X.; software, X.X.; validation, X.X., Y.Y. and Z.Z.; formal analysis, X.X.; investigation, X.X.; resources, X.X.; data curation, X.X.; writing---original draft preparation, X.X.; writing---review and editing, X.X.; visualization, X.X.; supervision, X.X.; project administration, X.X.; funding acquisition, Y.Y. All authors have read and agreed to the published version of the manuscript.'', please turn to the  \href{http://img.mdpi.org/data/contributor-role-instruction.pdf}{CRediT taxonomy} for the term explanation. Authorship must be limited to those who have contributed substantially to the work~reported.}

\funding{This research was funded by the IPLU grant on ``Quantifying Aliveness'' (Cambridge Planetary Science and Life in the Universe Research Grants Scheme 2021-22).\\\\
All data required to reproduce the results of this paper can be found at \url{https://doi.org/10.7910/DVN/FKKYY3}.} 

\acknowledgments{We thank Craig Walton and David Catling for helpful conversations about this work. We thank the anonymous referees and Shang-Min Tsai for constructive comments on the draft that helped to improve the work.}

\conflictsofinterest{The author declares no conflict of interest.} 

\appendixtitles{no} % Leave argument "no" if all appendix headings stay EMPTY (then no dot is printed after "Appendix A"). If the appendix sections contain a heading then change the argument to "yes".
\appendix
\section[\appendixname~\thesection]{Sensitivity Analysis}
\label{app:range}
\begin{figure}[H]
\centering
\includegraphics[width=0.8\textwidth]{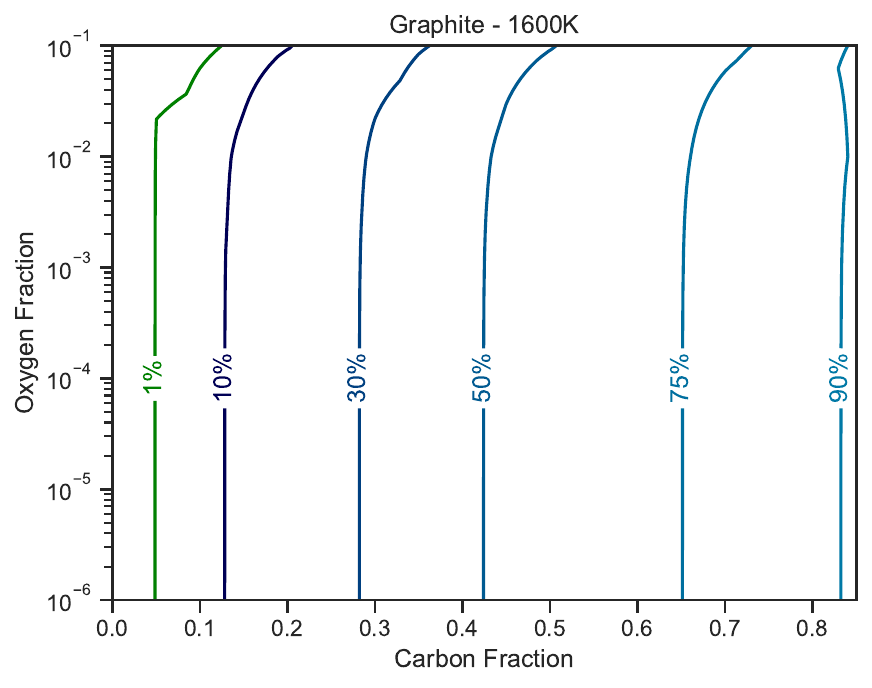}
\caption{Fraction of graphite in the mixture as a function of carbon (x-axis) and oxygen (y-axis) fractions for a magmatic gas held at $1600 \, {\rm ^{\circ}C}$ and $100 \, {\rm bar}$. The nitrogen fraction is set to $5\%$. The hydrogen fraction makes up the difference, if any.\label{fig:graphite-1600C}}
\end{figure}

\begin{figure}[H]
\centering
\includegraphics[width=0.8\textwidth]{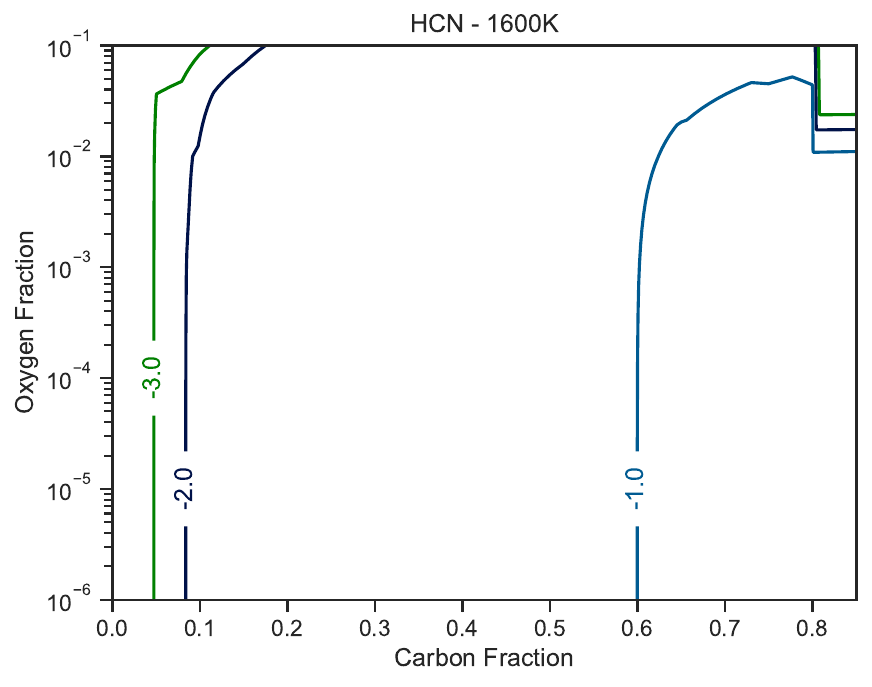}
\caption{Hydrogen cyanide concentrations, as $\log_{10} ([\ce{HCN}])$, as a function of carbon (x-axis) and oxygen (y-axis) fractions for a magmatic gas held at $1600 \, {\rm ^{\circ}C}$ and $100 \, {\rm bar}$. The nitrogen fraction is set to $5\%$. The hydrogen fraction makes up the difference, if any. \label{fig:HCN-1600C}}
\end{figure}

\begin{figure}[H]
\centering
\includegraphics[width=0.8\textwidth]{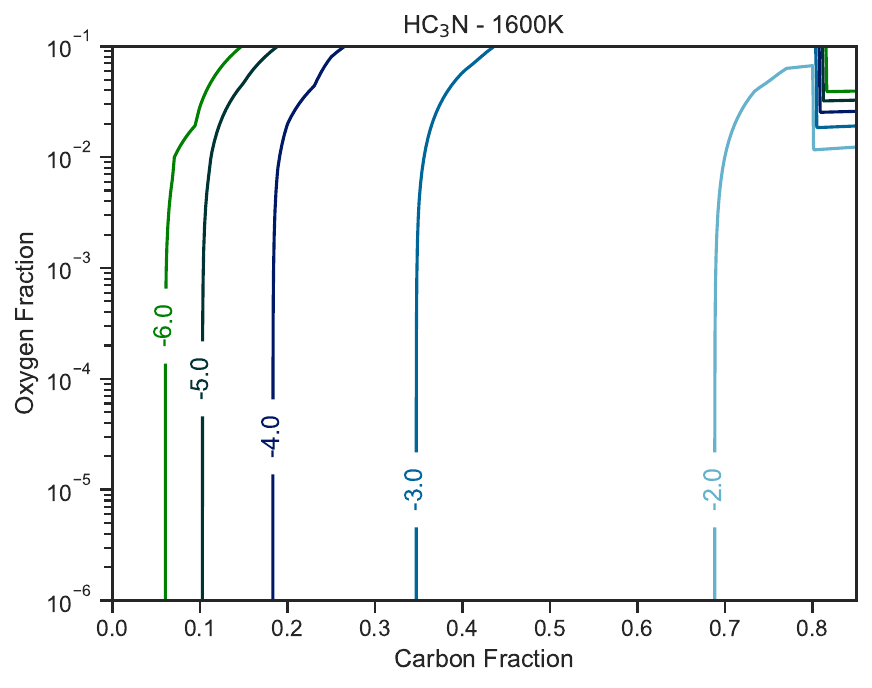}
\caption{Cyanoacetylene concentrations, as $\log_{10} ([\ce{HC_3N}])$, as a function of carbon (x-axis) and oxygen (y-axis) fractions for a magmatic gas held at $1600 \, {\rm ^{\circ}C}$ and $100 \, {\rm bar}$. The nitrogen fraction is set to $5\%$. The hydrogen fraction makes up the difference, if any.  \label{fig:HC3N-1600C}}
\end{figure}

\begin{figure}[H]
\centering
\includegraphics[width=0.8\textwidth]{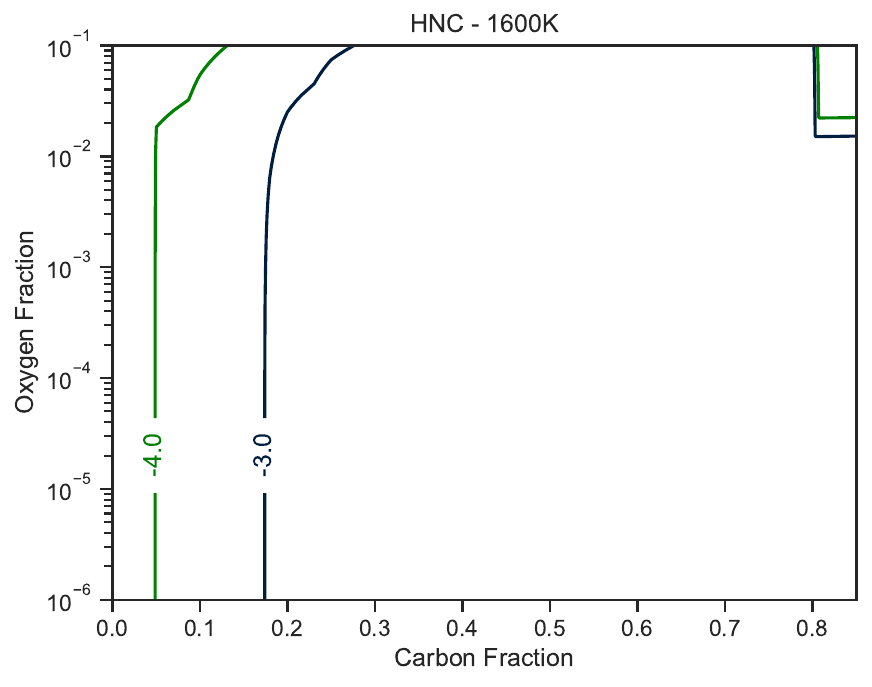}
\caption{Hydrogen isocyanide concentrations, as $\log_{10} ([\ce{HNC}])$, as a function of carbon (x-axis) and oxygen (y-axis) fractions for a magmatic gas held at $1600 \, {\rm ^{\circ}C}$ and $100 \, {\rm bar}$. The nitrogen fraction is set to $5\%$. The hydrogen fraction makes up the difference, if any. \label{fig:HNC-1600C}}
\end{figure}

\begin{figure}[H]
\centering
\includegraphics[width=0.8\textwidth]{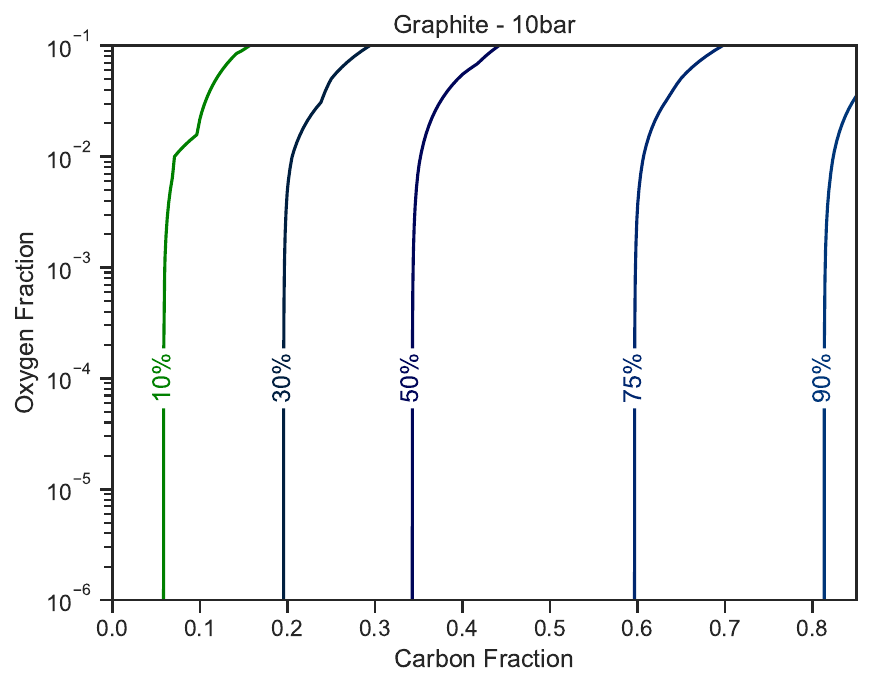}
\caption{Fraction of graphite in the mixture as a function of carbon (x-axis) and oxygen (y-axis) fractions for a magmatic gas held at $1600 \, {\rm ^{\circ}C}$ and $10 \, {\rm bar}$. The nitrogen fraction is set to $5\%$. The hydrogen fraction makes up the difference, if any.\label{fig:graphite-10bar}}
\end{figure}

\begin{figure}[H]
\centering
\includegraphics[width=0.8\textwidth]{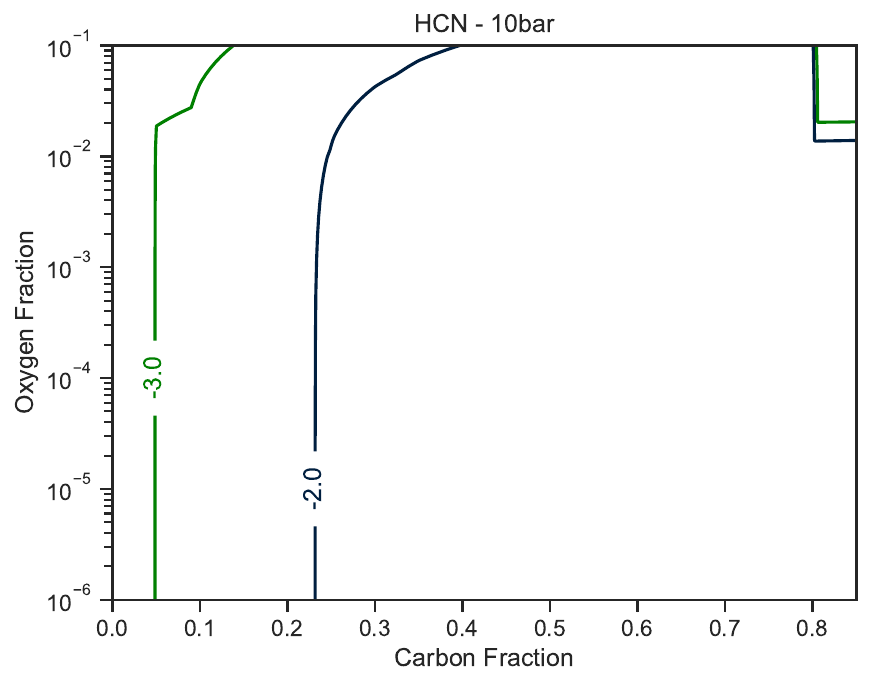}
\caption{Hydrogen cyanide concentrations, as $\log_{10} ([\ce{HCN}])$, as a function of carbon (x-axis) and oxygen (y-axis) fractions for a magmatic gas held at $1600 \, {\rm ^{\circ}C}$ and $10 \, {\rm bar}$. The nitrogen fraction is set to $5\%$. The hydrogen fraction makes up the difference, if any. \label{fig:HCN-10bar}}
\end{figure}

\begin{figure}[H]
\centering
\includegraphics[width=0.8\textwidth]{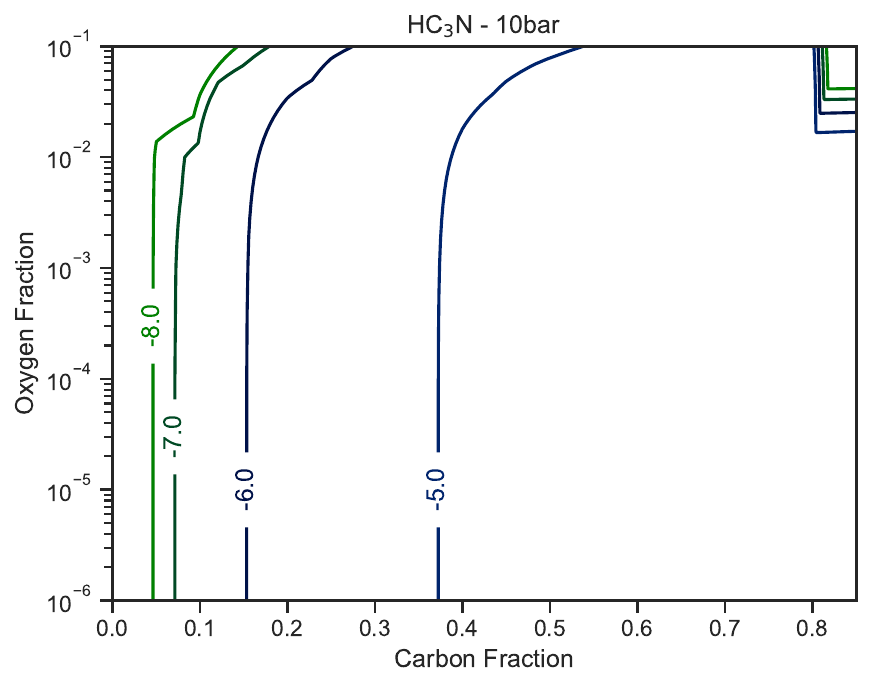}
\caption{Cyanoacetylene concentrations, as $\log_{10} ([\ce{HC_3N}])$, as a function of carbon (x-axis) and oxygen (y-axis) fractions for a magmatic gas held at $1600 \, {\rm ^{\circ}C}$ and $10 \, {\rm bar}$. The nitrogen fraction is set to $5\%$. The hydrogen fraction makes up the difference, if any.  \label{fig:HC3N-10bar}}
\end{figure}

\begin{figure}[H]
\centering
\includegraphics[width=0.8\textwidth]{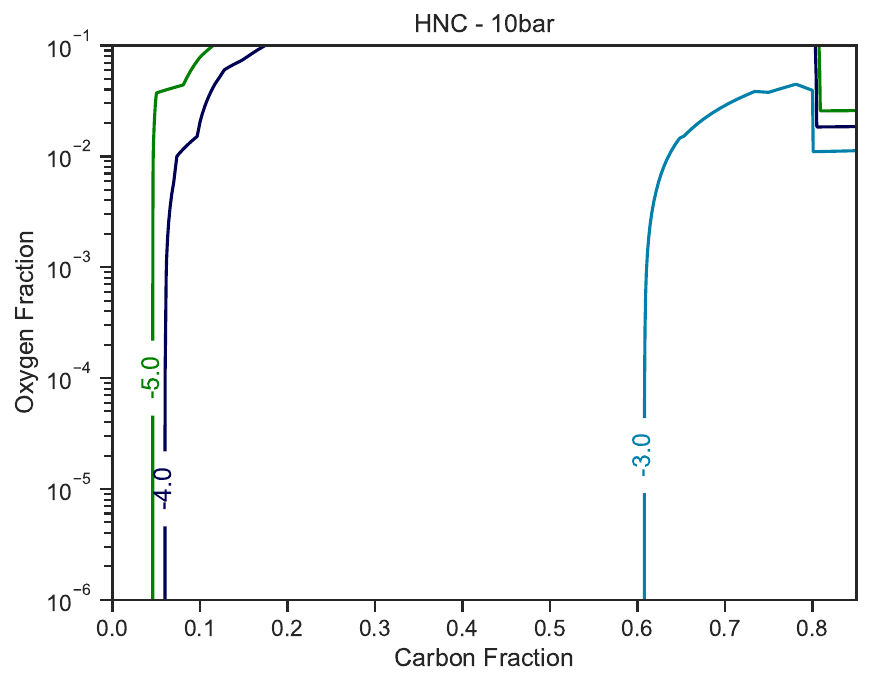}
\caption{Hydrogen isocyanide concentrations, as $\log_{10} ([\ce{HNC}])$, as a function of carbon (x-axis) and oxygen (y-axis) fractions for a magmatic gas held at $1600 \, {\rm ^{\circ}C}$ and $10 \, {\rm bar}$. The nitrogen fraction is set to $5\%$. The hydrogen fraction makes up the difference, if any. \label{fig:HNC-10bar}}
\end{figure}

\begin{figure}[H]
\centering
\includegraphics[width=0.8\textwidth]{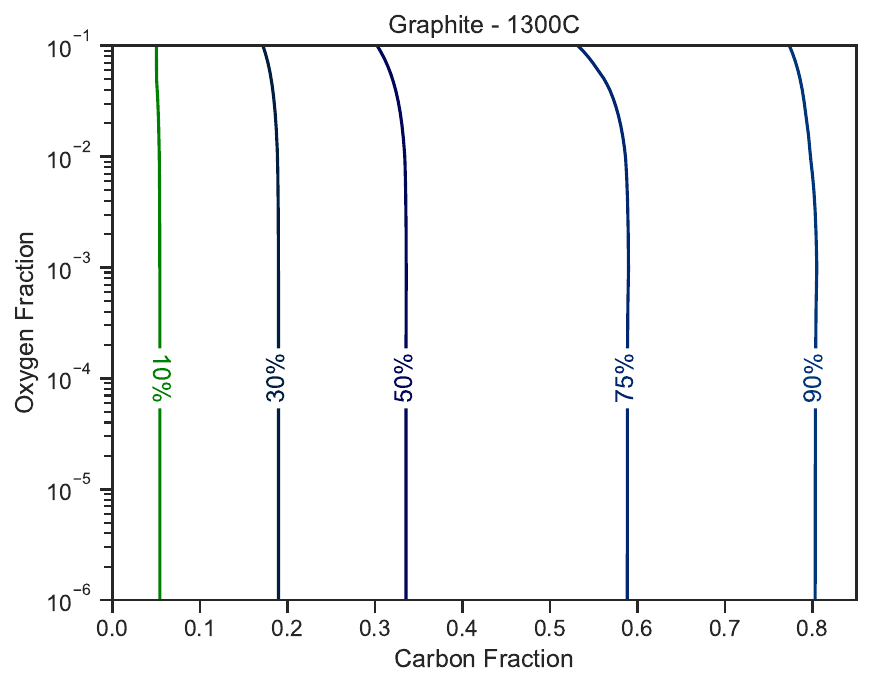}
\caption{Fraction of graphite in the mixture as a function of carbon (x-axis) and oxygen (y-axis) fractions for a magmatic gas held at $1300 \, {\rm ^{\circ}C}$ and $100 \, {\rm bar}$. The nitrogen fraction is set to $5\%$. The hydrogen fraction makes up the difference, if any.\label{fig:graphite-1300C}}
\end{figure}

\begin{figure}[H]
\centering
\includegraphics[width=0.8\textwidth]{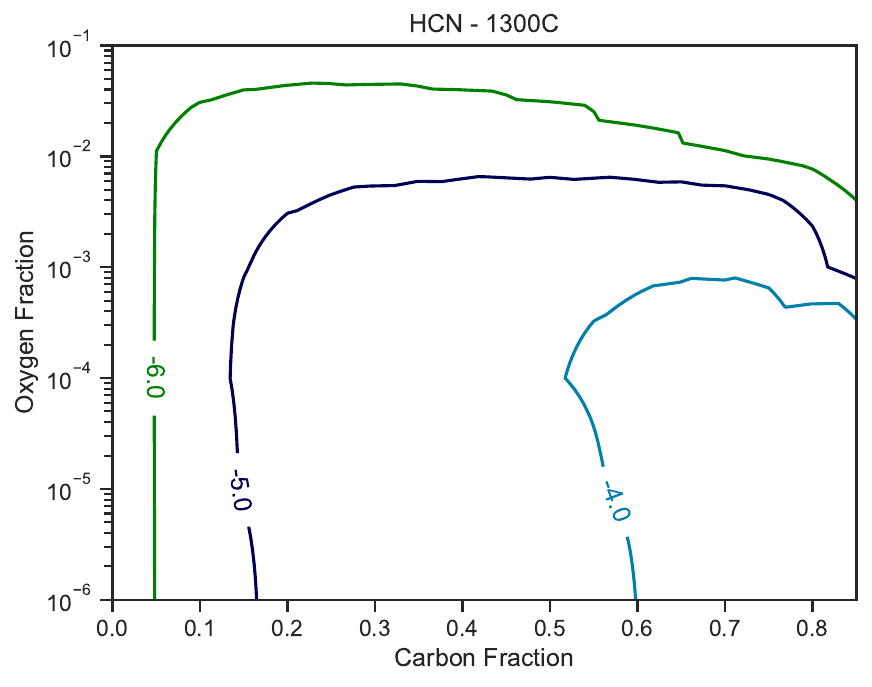}
\caption{Hydrogen cyanide concentrations, as $\log_{10} ([\ce{HCN}])$, as a function of carbon (x-axis) and oxygen (y-axis) fractions for a magmatic gas held at $1300 \, {\rm ^{\circ}C}$ and $100 \, {\rm bar}$. The nitrogen fraction is set to $5\%$. The hydrogen fraction makes up the difference, if any. \label{fig:HCN-1300C}}
\end{figure}

\begin{figure}[H]
\centering
\includegraphics[width=0.8\textwidth]{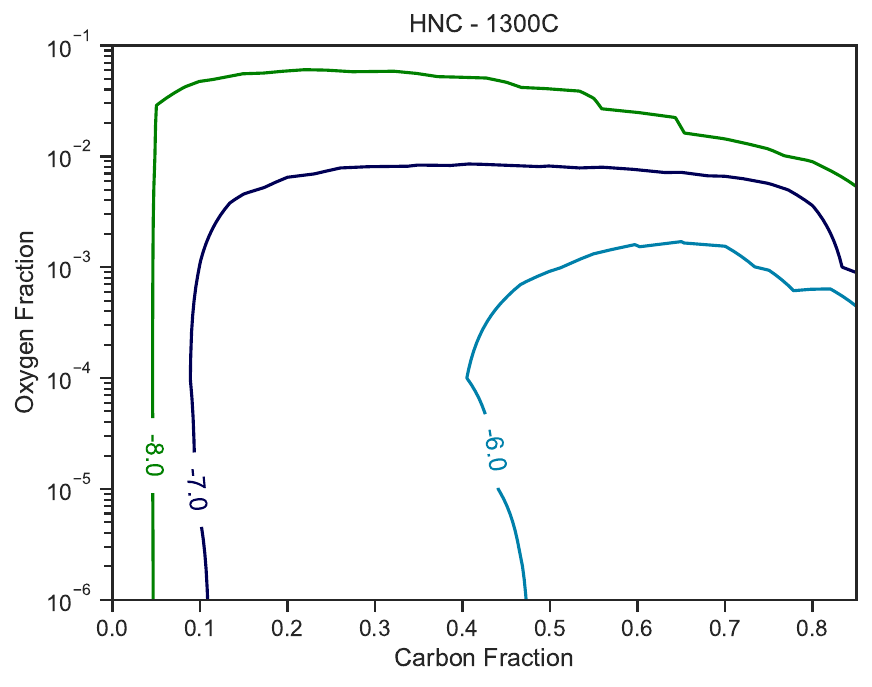}
\caption{Hydrogen isocyanide concentrations, as $\log_{10} ([\ce{HNC}])$, as a function of carbon (x-axis) and oxygen (y-axis) fractions for a magmatic gas held at $1300 \, {\rm ^{\circ}C}$ and $100 \, {\rm bar}$. The nitrogen fraction is set to $5\%$. The hydrogen fraction makes up the difference, if any. \label{fig:HNC-1300C}}
\end{figure}

%\subsection[\appendixname~\thesubsection]{}
Here we show the results of varying the gas-phase chemistry of the magma, by adjusting the elemental abundances of \ce{H}, \ce{C} and \ce{O} with a fixed \ce{N} at 5\%. These results are shown for a pressure of 100 bar and a temperature of $1600 \rm{\, ^{\circ}C}$: Figures \ref{fig:graphite-1600C} to \ref{fig:HNC-1600C}, as well as for a pressure of 10 bar and a temperature of $1600 \rm{\, ^{\circ}C}$: \ref{fig:graphite-10bar} to \ref{fig:HNC-10bar}, and for a pressure of 100 bar and a temperature of $1300 \rm{\, ^{\circ}C}$: \ref{fig:graphite-1300C} to \ref{fig:HNC-1300C}. For the last case, \ce{HC3N} is not shown because its abundance is below 1 ppm over the entire parameter space. \sed{Sensitivities based on these results are summarized in Table \ref{tab:range-summary}.} Full results, including all molecules predicted by the model and network, are given at \url{https://doi.org/10.7910/DVN/FKKYY3}. 

\sed{We find that in all cases that nitrile concentrations change linearly when varying the abundance of \ce{N}. The results when varying \ce{N} are therefore not plotted.}

\begin{table}[H] 
\caption{\sed{Sensitivity of species concentrations on temperature and pressure, expressed as ratios. $[\ce{X}_a/\ce{X}_b]$ is the ratio of $[\ce{X}]$ under the conditions specified by $a$ over $[\ce{X}]$ under the conditions specified by $b$. $a$ and $b$ are given in the top row. Unless otherwise stated, temperature is $1700 \, {\rm ^{\circ}C}$ and pressure is $100 \, {\rm bar}$.} \label{tab:range-summary}}
\begin{tabularx}{\textwidth}{ccc}
\toprule
\textbf{Species}	& \textbf{Temperature ($1300 \, {\rm ^{\circ}C}_a$/$1700 \, {\rm ^{\circ}C}_b$)}	& \textbf{Pressure $\;\;\;\;\;\;\;\;\;\;\;\;\;\;\;\;\;\;\;\;({\rm 10 \, bar_a/100 \, bar_b)}$}\\
\midrule
$[\ce{HCN}_a/\ce{HCN}_b]$	& $10^{-3}$	& $10^{-1}$\\
$[\ce{HC3N}_a/\ce{HC3N}_b]$	& $10^{-5}$			& $10^{-3}$\\
$[\ce{HNC}_a/\ce{HNC}_b]$	& $10^{-3}$ & $10^{-1}$\\
\bottomrule
\end{tabularx}
\end{table}

%100 bar; 1600 C

%10 bar, 1600 C

%100 bar; 1300 C

\reftitle{References}


\begin{thebibliography}{999}

\bibitem[{Islam} and {Powner}(2017)]{Islam2017}
{Islam}, S.; {Powner}, M.W.
\newblock {Prebiotic Systems Chemistry: Complexity Overcoming Clutter}.
\newblock {\em Chem} {\bf 2017}, {\em 2},~470--501.
\newblock {\url{https://doi.org/10.1016/j.chempr.2017.03.001}}.

\bibitem[Sutherland(2017)]{Sutherland2017}
Sutherland, J.D.
\newblock Opinion: Studies on the origin of life—the end of the beginning.
\newblock {\em Nature Reviews Chemistry} {\bf 2017}, {\em 1},~0012.

\bibitem[{Rimmer} et~al.(2018){Rimmer}, {Xu}, {Thompson}, {Gillen}, {Sutherland}, and {Queloz}]{Rimmer2018}
{Rimmer}, P.B.; {Xu}, J.; {Thompson}, S.J.; {Gillen}, E.; {Sutherland}, J.D.; {Queloz}, D.
\newblock {The origin of RNA precursors on exoplanets}.
\newblock {\em Science Advances} {\bf 2018}, {\em 4},~eaar3302,  \href{http://xxx.lanl.gov/abs/1808.02718}{{\normalfont [arXiv:astro-ph.EP/1808.02718]}}.
\newblock {\url{https://doi.org/10.1126/sciadv.aar3302}}.

\bibitem[Ruiz-Mirazo et~al.(2014)Ruiz-Mirazo, Briones, and de~la Escosura]{Ruiz2014}
Ruiz-Mirazo, K.; Briones, C.; de~la Escosura, A.
\newblock Prebiotic systems chemistry: new perspectives for the origins of life.
\newblock {\em Chemical reviews} {\bf 2014}, {\em 114},~285--366.

\bibitem[{Benner} et~al.(2019){Benner}, {Kim}, and {Biondi}]{Benner2019}
{Benner}, S.A.; {Kim}, H.J.; {Biondi}, E.
\newblock {Prebiotic Chemistry that Could Not Not Have Happened}.
\newblock {\em Life} {\bf 2019}, {\em 9},~84.
\newblock {\url{https://doi.org/10.3390/life9040084}}.

\bibitem[{Sasselov} et~al.(2020){Sasselov}, {Grotzinger}, and {Sutherland}]{Sasselov2020}
{Sasselov}, D.D.; {Grotzinger}, J.P.; {Sutherland}, J.D.
\newblock {The origin of life as a planetary phenomenon}.
\newblock {\em Science Advances} {\bf 2020}, {\em 6},~eaax3419.
\newblock {\url{https://doi.org/10.1126/sciadv.aax3419}}.

\bibitem[Green et~al.(2021)Green, Xu, and Sutherland]{Green2021}
Green, N.J.; Xu, J.; Sutherland, J.D.
\newblock Illuminating life’s origins: UV photochemistry in abiotic synthesis of biomolecules.
\newblock {\em Journal of the American Chemical Society} {\bf 2021}, {\em 143},~7219--7236.

\bibitem[Mariani et~al.(2018)Mariani, Russell, Javelle, and Sutherland]{Mariani2018}
Mariani, A.; Russell, D.A.; Javelle, T.; Sutherland, J.D.
\newblock A light-releasable potentially prebiotic nucleotide activating agent.
\newblock {\em Journal of the American Chemical Society} {\bf 2018}, {\em 140},~8657--8661.

\bibitem[{Liu} et~al.(2020){Liu}, {Wu}, {Xu}, {Bonfio}, {Russell}, and {Sutherland}]{Liu2020}
{Liu}, Z.; {Wu}, L.F.; {Xu}, J.; {Bonfio}, C.; {Russell}, D.A.; {Sutherland}, J.D.
\newblock {Harnessing chemical energy for the activation and joining of prebiotic building blocks}.
\newblock {\em Nature Chemistry} {\bf 2020}, {\em 12},~1--6.
\newblock {\url{https://doi.org/10.1038/s41557-020-00564-3}}.

\bibitem[Bonfio et~al.(2020)Bonfio, Russell, Green, Mariani, and Sutherland]{Bonfio2020}
Bonfio, C.; Russell, D.A.; Green, N.J.; Mariani, A.; Sutherland, J.D.
\newblock Activation chemistry drives the emergence of functionalised protocells.
\newblock {\em Chemical Science} {\bf 2020}, {\em 11},~10688--10697.

\bibitem[Wu et~al.(2021)Wu, Liu, and Sutherland]{Wu2021}
Wu, L.F.; Liu, Z.; Sutherland, J.D.
\newblock pH-Dependent peptide bond formation by the selective coupling of $\alpha$-amino acids in water.
\newblock {\em Chemical Communications} {\bf 2021}, {\em 57},~73--76.

\bibitem[{Benner}(2014)]{Benner2014}
{Benner}, S.A.
\newblock {Paradoxes in the Origin of Life}.
\newblock {\em Origins of Life and Evolution of the Biosphere} {\bf 2014}, {\em 44},~339--343.
\newblock {\url{https://doi.org/10.1007/s11084-014-9379-0}}.

\bibitem[{Patel} et~al.(2015){Patel}, {Percivalle}, {Ritson}, {Duffy}, and {Sutherland}]{Patel2015}
{Patel}, B.H.; {Percivalle}, C.; {Ritson}, D.J.; {Duffy}, C.D.; {Sutherland}, J.D.
\newblock {Common origins of RNA, protein and lipid precursors in a cyanosulfidic protometabolism}.
\newblock {\em Nature Chemistry} {\bf 2015}, {\em 7},~301--307.
\newblock {\url{https://doi.org/10.1038/nchem.2202}}.

\bibitem[{Sanchez} et~al.(1966){Sanchez}, {Ferris}, and {Orgel}]{Sanchez1966}
{Sanchez}, R.; {Ferris}, J.; {Orgel}, L.E.
\newblock {Conditions for Purine Synthesis: Did Prebiotic Synthesis Occur at Low Temperatures?}
\newblock {\em Science} {\bf 1966}, {\em 153},~72--73.
\newblock {\url{https://doi.org/10.1126/science.153.3731.72}}.

\bibitem[{Miyakawa} et~al.(2002){Miyakawa}, {James Cleaves}, and {Miller}]{Miyakawa2002}
{Miyakawa}, S.; {James Cleaves}, H.; {Miller}, S.L.
\newblock {The Cold Origin of Life: A. Implications Based On The Hydrolytic Stabilities Of Hydrogen Cyanide And Formamide}.
\newblock {\em Origins of Life and Evolution of the Biosphere} {\bf 2002}, {\em 32},~195--208.
\newblock {\url{https://doi.org/10.1023/A:1016514305984}}.

\bibitem[{Rimmer} et~al.(2021){Rimmer}, {Thompson}, {Xu}, {Russell}, {Green}, {Ritson}, {Sutherland}, and {Queloz}]{Rimmer2021Timescales}
{Rimmer}, P.B.; {Thompson}, S.J.; {Xu}, J.; {Russell}, D.A.; {Green}, N.J.; {Ritson}, D.J.; {Sutherland}, J.D.; {Queloz}, D.P.
\newblock {Timescales for Prebiotic Photochemistry Under Realistic Surface Ultraviolet Conditions}.
\newblock {\em Astrobiology} {\bf 2021}, {\em 21},~1099--1120.
\newblock {\url{https://doi.org/10.1089/ast.2020.2335}}.

\bibitem[{Mansy} and {Szostak}(2008)]{Mansy2008}
{Mansy}, S.S.; {Szostak}, J.W.
\newblock {Thermostability of model protocell membranes}.
\newblock {\em Proceedings of the National Academy of Science} {\bf 2008}, {\em 105},~13351--13355.
\newblock {\url{https://doi.org/10.1073/pnas.0805086105}}.

\bibitem[Mariani et~al.(2018)Mariani, Bonfio, Johnson, and Sutherland]{Mariani2018b}
Mariani, A.; Bonfio, C.; Johnson, C.M.; Sutherland, J.D.
\newblock pH-Driven RNA strand separation under prebiotically plausible conditions.
\newblock {\em Biochemistry} {\bf 2018}, {\em 57},~6382--6386.

\bibitem[{Hulshof} and {Ponnamperuma}(1976)]{Hulsof1976}
{Hulshof}, J.; {Ponnamperuma}, C.
\newblock {Prebiotic condensation reactions in an aqueous medium: A review of condensing agents}.
\newblock {\em Origins of Life} {\bf 1976}, {\em 7},~197--224.
\newblock {\url{https://doi.org/10.1007/BF00926938}}.

\bibitem[{Cousins} et~al.(2013){Cousins}, {Crawford}, {Carrivick}, {Gunn}, {Harris}, {Kee}, {Karlsson}, {Carmody}, {Cockell}, {Herschy}, and {Joy}]{Cousins2013}
{Cousins}, C.R.; {Crawford}, I.A.; {Carrivick}, J.L.; {Gunn}, M.; {Harris}, J.; {Kee}, T.P.; {Karlsson}, M.; {Carmody}, L.; {Cockell}, C.; {Herschy}, B.;  et~al.
\newblock {Glaciovolcanic hydrothermal environments in Iceland and implications for their detection on Mars}.
\newblock {\em Journal of Volcanology and Geothermal Research} {\bf 2013}, {\em 256},~61--77.
\newblock {\url{https://doi.org/10.1016/j.jvolgeores.2013.02.009}}.

\bibitem[{Ilanko} et~al.(2019){Ilanko}, {Fischer}, {Kyle}, {Curtis}, {Lee}, and {Sano}]{Ilanko2019}
{Ilanko}, T.; {Fischer}, T.P.; {Kyle}, P.; {Curtis}, A.; {Lee}, H.; {Sano}, Y.
\newblock {Modification of fumarolic gases by the ice-covered edifice of Erebus volcano, Antarctica}.
\newblock {\em Journal of Volcanology and Geothermal Research} {\bf 2019}, {\em 381},~119--139.
\newblock {\url{https://doi.org/10.1016/j.jvolgeores.2019.05.01710.31223/osf.io/jsc25}}.

\bibitem[{Toner} and {Catling}(2020)]{Toner2020}
{Toner}, J.D.; {Catling}, D.C.
\newblock {A carbonate-rich lake solution to the phosphate problem of the origin of life}.
\newblock {\em Proceedings of the National Academy of Science} {\bf 2020}, {\em 117},~883--888.
\newblock {\url{https://doi.org/10.1073/pnas.1916109117}}.

\bibitem[{Lin} et~al.(2005){Lin}, {Hall}, {Lippmann-Pipke}, {Ward}, {Sherwood Lollar}, {Deflaun}, {Rothmel}, {Moser}, {Gihring}, {Mislowack}, and {Onstott}]{Lin2005}
{Lin}, L.H.; {Hall}, J.; {Lippmann-Pipke}, J.; {Ward}, J.A.; {Sherwood Lollar}, B.; {Deflaun}, M.; {Rothmel}, R.; {Moser}, D.; {Gihring}, T.M.; {Mislowack}, B.;  et~al.
\newblock {Radiolytic H$_{2}$ in continental crust: Nuclear power for deep subsurface microbial communities}.
\newblock {\em Geochemistry, Geophysics, Geosystems} {\bf 2005}, {\em 6},~Q07003.
\newblock {\url{https://doi.org/10.1029/2004GC000907}}.

\bibitem[Shibuya et~al.(2015)Shibuya, Yoshizaki, Sato, Shimizu, Nakamura, Omori, Suzuki, Takai, Tsunakawa, and Maruyama]{Shibuya2015}
Shibuya, T.; Yoshizaki, M.; Sato, M.; Shimizu, K.; Nakamura, K.; Omori, S.; Suzuki, K.; Takai, K.; Tsunakawa, H.; Maruyama, S.
\newblock Hydrogen-rich hydrothermal environments in the Hadean ocean inferred from serpentinization of komatiites at 300 C and 500 bar.
\newblock {\em Progress in Earth and Planetary Science} {\bf 2015}, {\em 2},~1--11.

\bibitem[{Klein} et~al.(2013){Klein}, {Bach}, and {McCollom}]{Klein2013}
{Klein}, F.; {Bach}, W.; {McCollom}, T.M.
\newblock {Compositional controls on hydrogen generation during serpentinization of ultramafic rocks}.
\newblock {\em Lithos} {\bf 2013}, {\em 178},~55--69.
\newblock {\url{https://doi.org/10.1016/j.lithos.2013.03.008}}.

\bibitem[{Rimmer} and {Shorttle}(2019)]{Rimmer2019Hydro}
{Rimmer}, P.; {Shorttle}, O.
\newblock {Origin of Life's Building Blocks in Carbon- and Nitrogen-Rich Surface Hydrothermal Vents}.
\newblock {\em Life} {\bf 2019}, {\em 9},~12,  \href{http://xxx.lanl.gov/abs/1901.08542}{{\normalfont [arXiv:astro-ph.EP/1901.08542]}}.
\newblock {\url{https://doi.org/10.3390/life9010012}}.

\bibitem[{Gold} and {Soter}(1980)]{Gold1980}
{Gold}, T.; {Soter}, S.
\newblock {The Deep-Earth-Gas Hypothesis}.
\newblock {\em Scientific American} {\bf 1980}, {\em 242},~154--161.
\newblock {\url{https://doi.org/10.1038/scientificamerican0680-154}}.

\bibitem[Glasby(2006)]{Glasby2006}
Glasby, G.P.
\newblock Abiogenic origin of hydrocarbons: An historical overview.
\newblock {\em Resource Geology} {\bf 2006}, {\em 56},~83--96.

\bibitem[{Kim} et~al.(2016){Kim}, {Furukawa}, {Kakegawa}, {Bita}, {Scorei}, and {Benner}]{Kim2016}
{Kim}, H.J.; {Furukawa}, Y.; {Kakegawa}, T.; {Bita}, A.; {Scorei}, R.; {Benner}, S.A.
\newblock {Evaporite Borate-Containing Mineral Ensembles Make Phosphate Available and Regiospecifically Phosphorylate Ribonucleosides: Borate as a Multifaceted Problem Solver in Prebiotic Chemistry}.
\newblock {\em Angewandte Chemie} {\bf 2016}, {\em 128},~16048--16052.
\newblock {\url{https://doi.org/10.1002/ange.201608001}}.

\bibitem[Zahnle et~al.(2010)Zahnle, Schaefer, and Fegley]{Zahnle2010}
Zahnle, K.; Schaefer, L.; Fegley, B.
\newblock Earth’s earliest atmospheres.
\newblock {\em Cold Spring Harbor perspectives in biology} {\bf 2010}, {\em 2},~a004895.

\bibitem[{Zahnle} et~al.(2018){Zahnle}, {Gacesa}, and {Catling}]{Zahnle2018}
{Zahnle}, K.J.; {Gacesa}, M.; {Catling}, D.C.
\newblock {Strange messenger: A new history of hydrogen on Earth as told by xenon}.
\newblock In Proceedings of the AGU Fall Meeting Abstracts,  2018, Vol. 2018, pp. P44B--01.

\bibitem[{Ranjan} et~al.(2018){Ranjan}, {Todd}, {Sutherland}, and {Sasselov}]{Ranjan2018}
{Ranjan}, S.; {Todd}, Z.R.; {Sutherland}, J.D.; {Sasselov}, D.D.
\newblock {Sulfidic Anion Concentrations on Early Earth for Surficial Origins-of-Life Chemistry}.
\newblock {\em Astrobiology} {\bf 2018}, {\em 18},~1023--1040,  \href{http://xxx.lanl.gov/abs/1801.07725}{{\normalfont [arXiv:astro-ph.EP/1801.07725]}}.
\newblock {\url{https://doi.org/10.1089/ast.2017.1770}}.

\bibitem[Ranjan et~al.(2023)Ranjan, Abdelazim, Lozano, Mandal, Zhou, Kufner, Todd, Sahai, and Sasselov]{Ranjan2023}
Ranjan, S.; Abdelazim, K.; Lozano, G.G.; Mandal, S.; Zhou, C.Y.; Kufner, C.L.; Todd, Z.R.; Sahai, N.; Sasselov, D.D.
\newblock Geochemical and photochemical constraints on S [IV] concentrations in natural waters on prebiotic Earth.
\newblock {\em AGU Advances} {\bf 2023}, {\em 4},~e2023AV000926.

\bibitem[{Genda} et~al.(2017){Genda}, {Brasser}, and {Mojzsis}]{Genda2017}
{Genda}, H.; {Brasser}, R.; {Mojzsis}, S.J.
\newblock {The terrestrial late veneer from core disruption of a lunar-sized impactor}.
\newblock {\em Earth and Planetary Science Letters} {\bf 2017}, {\em 480},~25--32,  \href{http://xxx.lanl.gov/abs/1709.07554}{{\normalfont [arXiv:astro-ph.EP/1709.07554]}}.
\newblock {\url{https://doi.org/10.1016/j.epsl.2017.09.041}}.

\bibitem[{Itcovitz} et~al.(2022){Itcovitz}, {Rae}, {Citron}, {Stewart}, {Sinclair}, {Rimmer}, and {Shorttle}]{Itcovitz2022}
{Itcovitz}, J.P.; {Rae}, A.S.P.; {Citron}, R.I.; {Stewart}, S.T.; {Sinclair}, C.A.; {Rimmer}, P.B.; {Shorttle}, O.
\newblock {Reduced Atmospheres of Post-impact Worlds: The Early Earth}.
\newblock {\em Planetary Science Journal} {\bf 2022}, {\em 3},~115,  \href{http://xxx.lanl.gov/abs/2204.09946}{{\normalfont [arXiv:astro-ph.EP/2204.09946]}}.
\newblock {\url{https://doi.org/10.3847/PSJ/ac67a9}}.

\bibitem[{Zahnle} et~al.(2020){Zahnle}, {Lupu}, {Catling}, and {Wogan}]{Zahnle2020}
{Zahnle}, K.J.; {Lupu}, R.; {Catling}, D.C.; {Wogan}, N.
\newblock {Creation and Evolution of Impact-generated Reduced Atmospheres of Early Earth}.
\newblock {\em Planetary Science Journal} {\bf 2020}, {\em 1},~11,  \href{http://xxx.lanl.gov/abs/2001.00095}{{\normalfont [arXiv:astro-ph.EP/2001.00095]}}.
\newblock {\url{https://doi.org/10.3847/PSJ/ab7e2c}}.

\bibitem[{Wogan} et~al.(2023){Wogan}, {Catling}, {Zahnle}, and {Lupu}]{Wogan2023}
{Wogan}, N.F.; {Catling}, D.C.; {Zahnle}, K.J.; {Lupu}, R.
\newblock {Origin-of-life Molecules in the Atmosphere after Big Impacts on the Early Earth}.
\newblock {\em Planetary Science Journal} {\bf 2023}, {\em 4},~169,  \href{http://xxx.lanl.gov/abs/2307.09761}{{\normalfont [arXiv:astro-ph.EP/2307.09761]}}.
\newblock {\url{https://doi.org/10.3847/PSJ/aced83}}.

\bibitem[{Trainer} et~al.(2004){Trainer}, {Pavlov}, {Curtis}, {McKay}, {Worsnop}, {Delia}, {Toohey}, {Toon}, and {Tolbert}]{Trainer2004}
{Trainer}, M.G.; {Pavlov}, A.A.; {Curtis}, D.B.; {McKay}, C.P.; {Worsnop}, D.R.; {Delia}, A.E.; {Toohey}, D.W.; {Toon}, O.B.; {Tolbert}, M.A.
\newblock {Haze Aerosols in the Atmosphere of Early Earth: Manna from Heaven}.
\newblock {\em Astrobiology} {\bf 2004}, {\em 4},~409--419.
\newblock {\url{https://doi.org/10.1089/ast.2004.4.409}}.

\bibitem[{Arney} et~al.(){Arney}, {Meadows}, {Domagal-Goldman}, {Deming}, {Robinson}, {Tovar}, {Wolf}, and {Schwieterman}]{Arney2017}
{Arney}, G.N.; {Meadows}, V.S.; {Domagal-Goldman}, S.D.; {Deming}, D.; {Robinson}, T.D.; {Tovar}, G.; {Wolf}, E.T.; {Schwieterman}, E.
\newblock {Pale Orange Dots: The Impact of Organic Haze on the Habitability and Detectability of Earthlike Exoplanets}.
\newblock {\em Astrophysical Journal}, {\em 836},~49,  \href{http://xxx.lanl.gov/abs/1702.02994}{{\normalfont [arXiv:astro-ph.EP/1702.02994]}}.
\newblock {\url{https://doi.org/10.3847/1538-4357/836/1/49}}.

\bibitem[Benner et~al.(2020)Benner, Bell, Biondi, Brasser, Carell, Kim, Mojzsis, Omran, Pasek, and Trail]{Benner2020}
Benner, S.A.; Bell, E.A.; Biondi, E.; Brasser, R.; Carell, T.; Kim, H.J.; Mojzsis, S.J.; Omran, A.; Pasek, M.A.; Trail, D.
\newblock When did life likely emerge on Earth in an RNA-first process?
\newblock {\em ChemSystemsChem} {\bf 2020}, {\em 2},~e1900035.

\bibitem[{Ritson} et~al.(2020){Ritson}, {Mojzsis}, and {Sutherland}]{Ritson2020}
{Ritson}, D.J.; {Mojzsis}, S.J.; {Sutherland}, J.D.
\newblock {Supply of phosphate to early Earth by photogeochemistry after meteoritic weathering}.
\newblock {\em Nature Geoscience} {\bf 2020}, {\em 13},~344--348.
\newblock {\url{https://doi.org/10.1038/s41561-020-0556-7}}.

\bibitem[{Kadoya} et~al.(2020){Kadoya}, {Krissansen-Totton}, and {Catling}]{Kadoya2020}
{Kadoya}, S.; {Krissansen-Totton}, J.; {Catling}, D.C.
\newblock {Probable Cold and Alkaline Surface Environment of the Hadean Earth Caused by Impact Ejecta Weathering}.
\newblock {\em Geochemistry, Geophysics, Geosystems} {\bf 2020}, {\em 21},~e2019GC008734.
\newblock {\url{https://doi.org/10.1029/2019GC008734}}.

\bibitem[{Nisbet} et~al.(1993){Nisbet}, {Cheadle}, {Arndt}, and {Bickle}]{Nisbet1993}
{Nisbet}, E.G.; {Cheadle}, M.J.; {Arndt}, N.T.; {Bickle}, M.J.
\newblock {Constraining the potential temperature of the Archaean mantle: A review of the evidence from komatiites}.
\newblock {\em Lithos} {\bf 1993}, {\em 30},~291--307.
\newblock {\url{https://doi.org/10.1016/0024-4937(93)90042-B}}.

\bibitem[Burcat and Ruscic(2005)]{Burcat2005}
Burcat, A.; Ruscic, B.
\newblock Third millenium ideal gas and condensed phase thermochemical database for combustion (with update from active thermochemical tables).
\newblock Technical report, Argonne National Lab.(ANL), Argonne, IL (United States),  2005.

\bibitem[{Rimmer} and {Helling}(2016)]{Rimmer2016}
{Rimmer}, P.B.; {Helling}, C.
\newblock {A Chemical Kinetics Network for Lightning and Life in Planetary Atmospheres}.
\newblock {\em Astrophysical Journal Supplemental Series} {\bf 2016}, {\em 224},~9,  \href{http://xxx.lanl.gov/abs/1510.07052}{{\normalfont [arXiv:astro-ph.EP/1510.07052]}}.
\newblock {\url{https://doi.org/10.3847/0067-0049/224/1/9}}.

\bibitem[{Rimmer} et~al.(2021){Rimmer}, {Jordan}, {Constantinou}, {Woitke}, {Shorttle}, {Hobbs}, and {Paschodimas}]{Rimmer2021Venus}
{Rimmer}, P.B.; {Jordan}, S.; {Constantinou}, T.; {Woitke}, P.; {Shorttle}, O.; {Hobbs}, R.; {Paschodimas}, A.
\newblock {Hydroxide Salts in the Clouds of Venus: Their Effect on the Sulfur Cycle and Cloud Droplet pH}.
\newblock {\em Planetary Science Journal} {\bf 2021}, {\em 2},~133,  \href{http://xxx.lanl.gov/abs/2101.08582}{{\normalfont [arXiv:astro-ph.EP/2101.08582]}}.
\newblock {\url{https://doi.org/10.3847/PSJ/ac0156}}.

\bibitem[{Brewer} et~al.(1948){Brewer}, {Gilles}, and {Jenkins}]{Brewer1948}
{Brewer}, L.; {Gilles}, P.W.; {Jenkins}, F.A.
\newblock {The Vapor Pressure and Heat of Sublimation of Graphite}.
\newblock {\em Journal of Chemical Physics} {\bf 1948}, {\em 16},~797--807.
\newblock {\url{https://doi.org/10.1063/1.1746999}}.

\bibitem[{Joseph} et~al.(2002){Joseph}, {Sivakumar}, and {Manoravi}]{Joseph2002}
{Joseph}, M.; {Sivakumar}, N.; {Manoravi}, P.
\newblock {High temperature vapour pressure studies on graphite using laser pulse heating}.
\newblock {\em Carbon} {\bf 2002}, {\em 40},~2031--2034.
\newblock {\url{https://doi.org/10.1016/S0008-6223(02)00158-6}}.

\bibitem[{Stock} et~al.(2022){Stock}, {Kitzmann}, and {Patzer}]{Stock2022}
{Stock}, J.W.; {Kitzmann}, D.; {Patzer}, A.B.C.
\newblock {FASTCHEM 2 : an improved computer program to determine the gas-phase chemical equilibrium composition for arbitrary element distributions}.
\newblock {\em Monthly Notices of the Royal Astronomical Society} {\bf 2022}, {\em 517},~4070--4080,  \href{http://xxx.lanl.gov/abs/2206.08247}{{\normalfont [arXiv:astro-ph.EP/2206.08247]}}.
\newblock {\url{https://doi.org/10.1093/mnras/stac2623}}.

\bibitem[Knight et~al.(1985)Knight, Freeman, McEwan, Adams, and Smith]{knight1985selected}
Knight, J.; Freeman, C.; McEwan, M.; Adams, N.; Smith, D.
\newblock Selected-ion flow tube studies of HC3N.
\newblock {\em International journal of mass spectrometry and ion processes} {\bf 1985}, {\em 67},~317--330.

\bibitem[Coogan et~al.(2014)Coogan, Saunders, and Wilson]{coogan2014_chemgeol}
Coogan, L.; Saunders, A.; Wilson, R.
\newblock Aluminum-in-olivine thermometry of primitive basalts: Evidence of an anomalously hot mantle source for large igneous provinces.
\newblock {\em Chemical Geology} {\bf 2014}, {\em 368},~1--10.

\bibitem[{Palmer}(1967)]{Palmer1967}
{Palmer}, D.J.
\newblock {Reaction of Hydrogen with Graphite}.
\newblock {\em Nature} {\bf 1967}, {\em 215},~388--389.
\newblock {\url{https://doi.org/10.1038/215388a0}}.

\bibitem[{Wogan} et~al.(2020){Wogan}, {Krissansen-Totton}, and {Catling}]{Wogan2020}
{Wogan}, N.; {Krissansen-Totton}, J.; {Catling}, D.C.
\newblock {Abundant Atmospheric Methane from Volcanism on Terrestrial Planets Is Unlikely and Strengthens the Case for Methane as a Biosignature}.
\newblock {\em Planetary Science Journal} {\bf 2020}, {\em 1},~58,  \href{http://xxx.lanl.gov/abs/2009.07761}{{\normalfont [arXiv:astro-ph.EP/2009.07761]}}.
\newblock {\url{https://doi.org/10.3847/PSJ/abb99e}}.

\bibitem[Herzberg et~al.(2010)Herzberg, Condie, and Korenaga]{herzberg2010thermal}
Herzberg, C.; Condie, K.; Korenaga, J.
\newblock Thermal history of the Earth and its petrological expression.
\newblock {\em Earth and Planetary Science Letters} {\bf 2010}, {\em 292},~79--88.

\bibitem[{Takahashi} and {Scarfe}(1985)]{Takahashi1985}
{Takahashi}, E.; {Scarfe}, C.M.
\newblock {Melting of peridotite to 14 GPa and the genesis of komatiite}.
\newblock {\em Nature} {\bf 1985}, {\em 315},~566--568.
\newblock {\url{https://doi.org/10.1038/315566a0}}.

\bibitem[{Campbell} et~al.(1989){Campbell}, {Griffiths}, and {Hill}]{Campbell1989}
{Campbell}, I.H.; {Griffiths}, R.W.; {Hill}, R.I.
\newblock {Melting in an Archaean mantle plume: heads it's basalts, tails it's komatiites}.
\newblock {\em Nature} {\bf 1989}, {\em 339},~697--699.
\newblock {\url{https://doi.org/10.1038/339697a0}}.

\bibitem[{Nebel} et~al.(2014){Nebel}, {Campbell}, {Sossi}, and {Van Kranendonk}]{Nebel2014}
{Nebel}, O.; {Campbell}, I.H.; {Sossi}, P.A.; {Van Kranendonk}, M.J.
\newblock {Hafnium and iron isotopes in early Archean komatiites record a plume-driven convection cycle in the Hadean Earth}.
\newblock {\em Earth and Planetary Science Letters} {\bf 2014}, {\em 397},~111--120.
\newblock {\url{https://doi.org/10.1016/j.epsl.2014.04.028}}.

\bibitem[{Walton} et~al.(2023){Walton}, {Rigley}, {Lipp}, {Law}, {Suttle}, {Sch\"{o}nb\"{a}chler}, {Wyatt}, and {Shorttle}]{Walton2023}
{Walton}, C.R.; {Rigley}, J.K.; {Lipp}, A.; {Law}, R.; {Suttle}, M.D.; {Sch\"{o}nb\"{a}chler}, M.; {Wyatt}, M.; {Shorttle}, O.
\newblock {Cosmic dust fertilization of prebiotic chemistry on early Earth}.
\newblock {\em Nature Astronomy} {\bf 2023}, p. Accepted.

\bibitem[Rempfert et~al.(2023)Rempfert, Nothaft, Kraus, Asamoto, Evans, Spear, Matter, Kopf, and Templeton]{Rempfert2023}
Rempfert, K.R.; Nothaft, D.B.; Kraus, E.A.; Asamoto, C.K.; Evans, R.D.; Spear, J.R.; Matter, J.M.; Kopf, S.H.; Templeton, A.S.
\newblock Subsurface biogeochemical cycling of nitrogen in the actively serpentinizing Samail Ophiolite, Oman.
\newblock {\em Frontiers in Microbiology} {\bf 2023}, {\em 14},~1139633.

\bibitem[Pastorek et~al.(2020)Pastorek, Ferus, \v{C}uba, \v{S}r\'{a}mek, Ivanek, and Civi\v{s}]{Pastorek2020}
Pastorek, A.; Ferus, M.; \v{C}uba, V.; \v{S}r\'{a}mek, O.; Ivanek, O.; Civi\v{s}, S.
\newblock Primordial radioactivity and prebiotic chemical evolution: Effect of $\gamma$ radiation on formamide-based synthesis.
\newblock {\em The Journal of Physical Chemistry B} {\bf 2020}, {\em 124},~8951--8959.

\bibitem[{Barge} and {Price}(2022)]{Barge2022}
{Barge}, L.M.; {Price}, R.E.
\newblock {Diverse geochemical conditions for prebiotic chemistry in shallow-sea alkaline hydrothermal vents}.
\newblock {\em Nature Geoscience} {\bf 2022}, {\em 15},~976--981.
\newblock {\url{https://doi.org/10.1038/s41561-022-01067-1}}.

\end{thebibliography}
\end{document}